\documentstyle[graphicx,epsfig]{cjour}
\begin{document}



\authorrunninghead{P.~Grandcl\'ement et al.}
\titlerunninghead{Spectral method for scalar and vectorial Poisson equations}






\def \cont#1{{\mathcal C}^{#1}}

\title{A multi-domain spectral method for scalar and vectorial
Poisson equations with non-compact sources}

\author {P.~Grandcl\'ement, S.~Bonazzola, E.~Gourgoulhon and J.-A.~Marck}

\affil{D\'epartement d'Astrophysique Relativiste et de Cosmologie 
(UMR 8629 du C.N.R.S.), Observatoire de Paris, Section de Meudon, F-92195 
Meudon Cedex, France}
\email{Philippe.Grandclement@obspm.fr}

\abstract{We present a spectral method for solving elliptic equations
which arise in general relativity, namely three-dimensional scalar
Poisson equations, as well as generalized vectorial Poisson equations
of the type $\Delta \vec{N} + \lambda \vec{\nabla}(\vec{\nabla}\cdot
\vec{N}) = \vec{S}$ with $\lambda \not= -1$.  
The source can extend in all the Euclidean space
${\bf R}^3$, provided it decays at least as $r^{-3}$.  A multi-domain approach is used, along with
spherical coordinates $(r,\theta,\phi)$. In each domain, Chebyshev
polynomials (in $r$ or $1/r$) and spherical harmonics (in $\theta$ and
$\phi$) expansions are used.  If the source decays as 
$r^{-k}$ the
error of the numerical solution is shown to decrease at least as
$N^{-2(k-2)}$, where $N$ is the number of Chebyshev coefficients.
The error is even evanescent, i.e. decreases as $\exp(-N)$, if the
source does not contain any spherical harmonics of index $l\geq k -3$ (scalar
case) or $l\geq k-5$ (vectorial case).}


\begin{article}

\section {Introduction}

\subsection{Scalar and vectorial Poisson equations with non-compact sources}

The most common elliptic equations which occur in numerical 
relativity (for a recent review see \cite{SeideS99}) are
the scalar Poisson equation
\begin{equation} \label{e:poisson_scal}
  \Delta F = S \ , 
\end{equation}
and the (generalized) vector Poisson equation
\begin{equation} \label{e:poisson_vect}
\Delta \vec{N} + \lambda \vec{\nabla}(\vec{\nabla}\cdot \vec{N}) =
\vec{S} \ ,
\end{equation}
where $\lambda$ is a constant different from $-1$, typically $\lambda=1/3$. 
Contrary to the Newtonian case, where the source term $S$ contains
only the matter density, the sources of these equations have a non-compact
support. Moreover, 
the Einstein equations being non linear, the sources $S$ and $\vec{S}$ 
depends (usually quadratically) on the solutions $F$ and $\vec{N}$. This
means that equations (\ref{e:poisson_scal}) and (\ref{e:poisson_vect}) must
be solved by iterations. 

More precisely, within the 3+1 formalism (also called {\em Cauchy formulation})
of general relativity (see \cite{York79} for a review), the 10 Einstein
equations can be decomposed into a set of 6 second order evolution equations
and 4 constraint equations: a scalar one, the so-called {\em Hamiltonian 
constraint}, and a vectorial one, the so-called {\em momentum constraint}
(see \cite{ChoquIY00} for an extensive discussion of the constraints 
equations).
The PDE type (i.e. hyperbolic, parabolic or
elliptic) of these equations depend on the 
coordinates chosen to describe the space-time manifold. 
Let us recall that within the 3+1 formalism, the space-time is foliated
in a family of space-like slices $\Sigma_t$, labeled by the time
coordinate $t$. The space-time 4-metric is then entirely described by
the induced 3-metric $\gamma_{ij}$ of the hypersurfaces $\Sigma_t$ along
with the extrinsic curvature tensor $K_{ij}$ of $\Sigma_t$. 

In this context, 
a typical example for Eq.~(\ref{e:poisson_scal}) is the equation for the
lapse function for the choice of time coordinate corresponding to 
a {\em maximal slicing} of space-time\footnote{This Poisson equation for 
the lapse function reduces to the usual Poisson equation for the gravitational
potential at the Newtonian limit} (see e.g. \cite{SmarrY78}). 
Another example is provided by York treatment of the initial-value 
problem of general relativity \cite{York73}, according to which the
Hamiltonian constraint equation results in an elliptic equation of the
type (\ref{e:poisson_scal})
for the conformal factor of the spatial metric $\gamma_{ij}$, 
with a term $F^{-7}$ in $S$. 

Regarding the vector Poisson equation (\ref{e:poisson_vect}), it
also appears in York formulation of the initial-value problem for 
the vector which enters in the longitudinal part of the transverse-traceless
decomposition of the extrinsic curvature tensor $K_{ij}$. Indeed the
momentum constraint determines the longitudinal part of $K_{ij}$ 
according to the equation\footnote{Einstein convention of summation
on repeated indices is used}
\begin{equation}
	\nabla_j K^{ij} = 8\pi J^i \ , 
\end{equation}
where $\nabla_j$ is the covariant derivative associated with the 3-metric
$\gamma_{ij}$, $J^i$ is the matter momentum density, 
and maximal slicing is assumed ($K_i^i=0$). 
More generally, the vector Poisson equation (\ref{e:poisson_vect})
with $\lambda=1/3$ occurs each time one has to perform the 
transverse-traceless decomposition of a symmetric tensor field $T^{ij}$
defined on a Riemannian three-manifold with metric $\gamma_{ij}$.
Following \cite{York73,York74}, this decomposition writes
\begin{equation}
\label{e:decomp}
T^{ij} = T^{ij}_{\rm TT} + \left(LY\right)^{ij} + \frac{1}{3} T \gamma^{ij} \ , 
\end{equation}
where $T = \gamma_{kl} T^{kl}$. $T_{\rm TT}^{ij}$ is the transverse-traceless part, 
$\left(LY\right)^{ij}$ the longitudinal trace-free one and 
$\frac{1}{3} T \gamma^{ij}$ the trace part.
The longitudinal part is expressible in term of a vector $Y^i$, by means of the 
conformal Killing operator~:
\begin{equation}
\left(LY\right)^{ij} = \nabla^i Y^j + \nabla^j Y^i 
	- \frac{2}{3}\gamma^{ij}\nabla_k Y^k \ . 
\end{equation}
Performing the decomposition reduces to the finding of the vector field 
$\vec{Y}$. Considering the divergence of Eq. (\ref{e:decomp}), $\vec{Y}$ appears to 
be the solution of the equation
\begin{equation}
\Delta Y^i + \frac{1}{3} \nabla^i \left(\nabla_j Y^j \right) = 
\nabla_j\left(T^{ij}-\frac{1}{3}T\gamma^{ij}\right)-R_j^i Y^j \ , 
\end{equation}
where $R_j^i$ is the Ricci tensor associated with the metric $\gamma_{ij}$. 
This is a vectorial Poisson equation 
of type (\ref{e:poisson_vect}) with $\lambda = \frac{1}{3}$ (involving the 
so-called conformal Laplace operator). Let us mention that, in the 
general case, it must be solved by iteration for $\vec{Y}$ is present in 
the source term.

Another example for the vectorial Poisson equation (\ref{e:poisson_vect})
is provided by the so-called {\em minimal distortion} \cite{SmarrY78}
choice of coordinates in the spatial hypersurfaces $\Sigma_t$. 
The unknown vector $\vec{N}$ is in this case the {\em shift vector}
which defines the propagation of the spatial coordinates $x^i$ from one
slice $\Sigma_t$ to the next one $\Sigma_{t+dt}$. 
It is this vectorial Poisson equation, which is a special form
of Eq.~(\ref{e:poisson_vect}) with $\lambda = \frac{1}{3}$, that 
originally motivated our study of this subject. Let us mention that the 
conformal Killing operator and the associated vectorial Poisson equation also 
appear in the ``thin-sandwich'' formulation, where the spatial geometry is 
given on two close hypersurfaces (see \cite{ChoquIY00, York98} for more 
details).

\subsection{Treatment by means of spectral methods}

Solving elliptic equations is often considered as a CPU-time consuming
task. Spectral methods \cite{CanutHQZ88, GottlO77} seems attractive in this
respect because they provide accurate results with reasonable sampling, 
as compared with finite difference methods for example.
We refer the interested reader to Refs.~\cite{BonazGM99b,Fraue99} for a review 
of the use of spectral methods in relativistic astrophysics. 
Let us simply mention here that 
our group has previously developed a spectral method, using Chebyshev
polynomials and spherical harmonics to solve three-dimensional scalar
Poisson equations with a compact source \cite{BonazM90}. 
However as recalled above, the elliptic equations which arise from numerical
relativity have non-compact sources. This means in particular that the
infinity is the only location to impose exact boundary conditions 
(flat space-time). In order to tackle this, we have introduced a 
multi-domain approach \cite{BonazGM98} 
within which the last domain extends up to infinity, thanks to some
compactification. 
This approach has another nice feature, for it is avoiding 
Gibbs phenomena: a physical discontinuity can be located at the
boundary between two domains, so that all the considered fields are
smooth in each domains. 

In this article, we extend the single-domain spectral method for the
scalar Poisson equation (\ref{e:poisson_scal}) 
presented in \cite{BonazM90} to the multi-domain case, which enables
in particular to treat non-compact sources provided they decay
at least as $r^{-3}$ when $r\rightarrow\infty$. Based on this scalar 
Poisson solver, we treat the generalized vectorial Poisson equation
(\ref{e:poisson_vect}). We consider three different schemes proposed in 
the literature to reduce the resolution of (\ref{e:poisson_vect}) to 4
scalar Poisson equations, namely the schemes of  
Bowen \& York \cite{BowenY80}, Oohara \& Nakamura  
\cite{OoharN97} and Oohara, Nakamura \& Shibata \cite{OoharNS97}. 
These schemes have been originally implemented on
finite (single) domains and with finite difference methods. 
We study here their applicability to 
infinite domains and spectral methods. 

The solvers presented in this work deal with three-dimensional
flat spaces where $\vec{\nabla}$ denotes the ordinary derivation. More general 
cases (i.e. Laplacian operator associated with a curved metric)
can be solved by iteration. In all the following we will assume that 
there exists a unique solution of both the scalar and vectorial equation 
that is $\cont{\infty}$ by parts, $\cont{1}$ everywhere and that is going 
to zero at infinity. For known results about the existence and uniqueness of 
solution of partial derivative systems see for example \cite{DautrL87}.

This paper is organized as follows. In Sec. \ref{scalar_eq} we present 
the numerical scheme used to solve the scalar Poisson equation with our 
multi-domain spectral method. This scheme is tested in Sec. \ref{secconv}
using comparison with analytical solutions of various behaviors. This 
study leads us to establish the convergence properties of the algorithm. Sec.
\ref{secvect} is devoted to the study the three different schemes 
mentioned above to solve
the vectorial Poisson equation (\ref{e:poisson_vect}). 
As for the scalar Poisson equation, the implemented schemes are tested in 
Sec. \ref{secvectconv} and their convergence properties exhibited. In
Sec. \ref{sec_devel} we give to the reader some indications about some extensions 
of this work that have been successfully conducted or under work.

\section{Scalar Poisson equation} \label{scalar_eq}

\subsection {Spectral expansions}

As described in previous articles \cite{BonazGM99b,BonazM90}, spherical 
coordinates $\left(r,\theta,\phi\right)$ are used; the fields are 
expanded in spherical harmonics $Y_l^m\left(\theta,\phi\right)$ and a 
Chebyshev expansion 
is performed with respect to the $r$ 
coordinate. Doing so the resolution of the scalar Poisson equation is reduced to 
find, for each couple $\left(l,m\right)$, the solution of

\begin{equation}
\label {poisson1d}
\frac{{\mathrm d}^2 f }{{\mathrm d} r^2}+\frac{2}{r}\frac{{\mathrm d} f}
{{\mathrm d} r}
-\frac{l\left(l+1\right)}{r^2} f = s\left(r\right)
\end{equation}
where $f$ and $s$ are functions of $r$ solely, being respectively the coefficients
 of $Y_l^m$ in the solution $F$ and in the source $S$.

$f$ and $s$ are expanded in Chebyshev polynomials (hereafter referred as $T_i$
for the polynomial of order $i$) so that the inversion of the 
operator on the left-hand-side of Eq. (\ref{poisson1d}) is reduced to a matrix 
inversion.

As recalled above, 
the present work improves that presented in \cite{BonazM90} for we are 
allowing a source that is not compactly supported. To take care of this, we 
will divide space in three type of domains, following \cite{BonazGM98}
\begin{itemize}
\item  One {\em kernel}, a sphere centered at the origin and being the only domain
considered in \cite{BonazM90}. In such a domain $r$ 
is given by $r = \alpha x$, where $x \in \left[0,1\right]$, with $\alpha > 0$.
The functions are expanded in Chebyshev polynomials in $x$ with a definite parity 
to ensure regularity at the origin : only even (resp. odd) polynomials are 
involved for $l$ even (resp. odd).

\item An arbitrary number, including zero, of {\em shells}, domains where $r=\alpha x + \beta$, 
$x \in \left[-1,1\right]$. We have the following conditions~:
$\alpha > 0$ and $\beta \geq \alpha$, so that $r$ is increasing with $x$ and never 
equal to zero. In the shells, the functions are expanded in usual Chebyshev 
polynomials, with no parity requirement.

\item One external domain, extending to infinity, where $r$ is given by 
$u= r^{-1}= \alpha (x-1)$, $\alpha$ being negative and $x \in \left[-1,1\right]$.
Once more the functions are given as a sum of Chebyshev polynomials in $x$.
\end{itemize}

\subsection{The matrices} \label{matrix}

Before doing any operator-inversion, one has to take care about singularities at 
the origin and at infinity. For example, because of division by $r^2$, 
the solution of the equation, must be decreasing as $r^2$ at the origin to 
be associated with a non-singular source. We choose to treat that by subtracting
 finite parts of the solution at the point of singularity.

Before describing that more precisely, let us mention another method for solving
that problem, presented in \cite{BonazM90}. In this Reference, the functions are expanded 
on a new set of basis-functions, that verify individually the regularity 
conditions (Galerkin basis) . For example, $T_{i+2}+T_{i}$ is used in the kernel, making all 
the basis-functions decrease as $r^{2}$ at the origin.
\begin{itemize}
\item In the kernel, we have to take care of a singularity at the origin due to 
the division
 by $r^2$. To avoid this we construct an operator without the finite part of $f$
at $x=0$. 
Thus the operator is, expressed in terms of $x$,

\begin{equation}
Af = 
\frac{{\mathrm d}^2 f}{{\mathrm d}x^2}+
\frac{2}{x}\left(\frac{{\mathrm d}f}{{\mathrm d}x}
-\frac{{\mathrm d}f}{{\mathrm d}x}\left(0\right)\right)
-\frac{l\left(l+1\right)}{x^2}\left(f-f\left(0\right)
-x\frac{{\mathrm d}f}{{\mathrm d}x}\left(0\right)\right),
\end{equation}
the source $s$ being multiplied by $\alpha^2$.

\item In the shells there is no singularity, so we can multiply the 
source by $\displaystyle\frac{r^2}{\alpha^2}$ and invert the following operator

\begin{equation}
Af = \left(x+\frac{\beta}{\alpha}\right)^2\frac{{\mathrm d}^2f}{{\mathrm d}x^2}
+2\left(x+\frac{\beta}{\alpha}\right)\frac{{\mathrm d}f}{{\mathrm d}x}
-l\left(l+1\right)f.
\end{equation}

\item In the external domain Eq. (\ref{poisson1d}), once re-written in terms of 
$u = \frac{1}{r}$, becomes
\begin{equation}
u^4\left(\frac{{\mathrm d}^2 f}{{\mathrm d}u^2} - \frac{l\left(l+1\right)}{u^2}f\right) = s.
\end{equation}

We consider the three following possibilities.

\begin{itemize}
\item
First multiplying the source by $r^4$ in the 
external domain, a singularity occurs at $r=\infty$, that is $x=1$. We handle it 
like in the kernel, by subtraction of the finite part of $f$ in $1$, 
and we use the following operator
\begin{equation}
Af=\frac{{\mathrm d}^2f}{{\mathrm d}x^2} 
-\frac{l\left(l+1\right)}{\left(x-1\right)^2}\left(f-f\left(1\right)
-\left(x-1\right)\frac{{\mathrm d}f}{{\mathrm d}x}\left(1\right)\right).
\end{equation}

\item
If the source is multiplied by $r^3$, a singularity occurs $r=\infty$, that is 
$x=1$ and is handled by the finite part method, so that the operator becomes~
\begin{equation}
Af=\left(x-1\right)\frac{{\mathrm d}^2f}{{\mathrm d}x^2} 
-\frac{l\left(l+1\right)}{\left(x-1\right)}\left(f-f\left(1\right)\right).
\end{equation}

\item
If the source is only multiplied by $r^2$, we invert the non-singular 
operator 
\begin{equation}
Af = \left(x-1\right)^2\frac{{\mathrm d}^2 f}{{\mathrm d}x^2}-l\left(l+1\right)f.
\end{equation}
\end{itemize}

In all cases we have to multiply $s$ by $\alpha^2$. Let us stress out that
those three operators are not fully equivalent in actual physical calculations 
based on iterative schemes. The effective source (i.e. $r^k S$) being given, the solution will have 
less high-frequency terms (Chebyshev polynomials of high-order), if the 
number $k$ is high. Those high-frequency terms may cause instabilities 
in an iterative procedure, so we always use the $r^4 S$ scheme except for
a source decreasing like $r^3 S$ at infinity.
\end{itemize}

As an illustration, here is the matrix constructed in the kernel, with $l=2$ and 
$9$ coefficients in $r$ (Chebyshev polynomials $T_0$, $T_2$, ..., $T_{16}$)
$$
\pmatrix{
0&0&56&96&304&480&936&1344&2144&\cr
0&0&56&240&472&1056&1656&2832&3992\cr
0&0&0&144&432&848&1632&2512&3984\cr
0&0&0&0&264&688&1320&2336&3528\cr
0&0&0&0&0&416&1008&1888&3168\cr
0&0&0&0&0&0&600&1392&2552\cr
0&0&0&0&0&0&0&816&1840\cr
0&0&0&0&0&0&0&0&1064\cr
0&0&0&0&0&0&0&0&0\cr}.
$$

\subsection {The banded-matrices}
The constructed matrices are not suitable for numerical purpose. The inversion 
would be much more rapid and much more efficient if we could work on band-
matrix instead of triangular ones. The operators being second order 
operators on a set of orthogonal functions, there must exist a linear 
combination of the lines so that the matrices are reduced to banded-ones 
(see \cite{Tucke89}).

\noindent We exhibit here the combination we used in each domain~:

\begin{itemize}
\item In the kernel, the Chebyshev polynomials are either odd or even,
 depending on the parity of $l$. 
The combination is independent of the actual value of $l$ except for its parity.

When the Chebyshev polynomials are even we use~:
\begin{eqnarray}
\overline{L}_i &=& \left(1+\delta_0^i\right)L_i-L_{i+2}
\phantom{00} \left({\mathrm for} \phantom{0} 0\leq i \leq N-3 \right)\\
\tilde{L}_i &=& \overline{L}_i-\overline{L}_{i+2}
\phantom{00} \left({\mathrm for} \phantom{0} 0\leq i \leq N-5 \right)\\
\dot{L}_i&=&\tilde{L}_i-\tilde{L}_{i+1}
\phantom{00} \left({\mathrm for} \phantom{0} 0\leq i \leq N-5 \right)
\end{eqnarray}

and when they are odd~:
\begin{eqnarray}
\overline{L}_i &=& L_i-L_{i+2}
\phantom{00} \left({\mathrm for} \phantom{0} 0\leq i \leq N-3 \right)\\
\tilde{L}_i &=& \overline{L}_i-\overline{L}_{i+2}
\phantom{00} \left({\mathrm for} \phantom{0} 0\leq i \leq N-5 \right)\\
\dot{L}_i&=&\tilde{L}_i-\tilde{L}_{i+1}
\phantom{00} \left({\mathrm for} \phantom{0} 0\leq i \leq N-5 \right)
\end{eqnarray}
where $L_i$ denotes the line number $i$ and $N$ is the number of 
Chebyshev polynomials involved in the expansion.

In both cases the resulting matrix is a 4-band one.

\item In the shells, the basis of decomposition contains all the Chebyshev 
polynomials. The combination is the following~:
\begin{eqnarray}
\overline{L}_i &=& \frac{\left(1+\delta_0^i\right)L_i-L_{i+2}}{i+1}
\phantom{00} \left({\mathrm for} \phantom{0} 0\leq i \leq N-3 \right)\\
\dot{L}_i &=& \overline{L}_i-\overline{L}_{i+2}
\phantom{00} \left({\mathrm for} \phantom{0} 0\leq i \leq N-5 \right)
\end{eqnarray}
The resulting matrix is a 5-band one.

\item In the external domain, the combination depends of the type of the 
constructed operator.
\begin{itemize}
\item If the source is multiplied by $r^4$ the combination is~:
\begin{eqnarray}
\overline{L}_i &=& \left(1+\delta_0^i\right)L_i-L_{i+2}
\phantom{00} \left({\mathrm for} \phantom{0} 0\leq i \leq N-3 \right)\\
\tilde{L}_i &=& \overline{L}_i-\overline{L}_{i+2}
\phantom{00} \left({\mathrm for} \phantom{0} 0\leq i \leq N-5 \right)\\
L'_i&=&\tilde{L}_i-\tilde{L}_{i+1}
\phantom{00} \left({\mathrm for} \phantom{0} 0\leq i \leq N-5 \right)\\
\dot{L}_i&=&L'_i-L'_{i+2}
\phantom{00} \left({\mathrm for} \phantom{0} 0\leq i \leq N-5 \right)
\end{eqnarray}
The resulting matrix is a 4-band one.

\item If the source is multiplied by $r^3$ the combination is~:
\begin{eqnarray}
\overline{L}_i &=& \left(1+\delta_0^i\right)L_i-L_{i+2}
\phantom{00} \left({\mathrm for} \phantom{0} 0\leq i \leq N-3 \right)\\
\tilde{L}_i &=& \overline{L}_i-\overline{L}_{i+2}
\phantom{00} \left({\mathrm for} \phantom{0} 0\leq i \leq N-5 \right)\\
\dot{L}_i&=&\tilde{L}_i+\tilde{L}_{i+1}
\phantom{00} \left({\mathrm for} \phantom{0} 0\leq i \leq N-5 \right)
\end{eqnarray}
The resulting matrix is a 4-band one.

\item If the source is only multiplied by $r^2$, the combination is the same 
than the one used in the kernel for even polynomials. Then, the resulting 
matrix is a 6-band one.
\end{itemize}
\end{itemize}

Of course to maintain the solution, the same linear combination is performed 
on the coefficients of $s$.

The banded matrix associated with the one presented above (in the kernel with
$l=2$ and $N=9$) is~:
$$
\pmatrix{
0&0&56&-336&-200&0&0&0&0\cr
0&0&56&96&-488&336&0&0&0\cr
0&0&0&144&168&-672&-504&0&0\cr
0&0&0&0&264&272&-888&-704&0\cr
0&0&0&0&0&416&408&-1136&-2000\cr
0&0&0&0&0&0&600&1392&1488\cr
0&0&0&0&0&0&0&816&1840\cr
0&0&0&0&0&0&0&0&1064\cr
0&0&0&0&0&0&0&0&0\cr}
$$

\subsection {Homogeneous solutions}

Due to the presence of homogeneous solutions, the banded-matrices are not 
invertible. The operator given by Eq. (\ref{poisson1d}) has two homogeneous 
solutions which are $r^l$ and $r^{-\left(l+1\right)}$. Those 
functions are eigenvectors of the matrix with the eigenvalue $0$. In the 
kernel and the external domain, the use of the finite part of the solution 
can sometimes introduce other homogeneous solutions.

\noindent Let us summarize the number of such eigenvectors in each case~:

\begin{itemize}
\item In the kernel the solution 
in $r^{-\left(l+1\right)}$ is singular for $r=0$ and so is not taken into account.
We have one additional homogeneous solution, arising from the 
finite part~: $T_0$ for $l$ even and $T_1$ for $l$ odd.

The parity of the Chebyshev polynomials is the same than that of $l$ so 
the eigenvectors are~:
\begin{itemize}
\item $T_0$ only for $l=0$.
\item $T_1$ only for $l=1$.
\item $r^l$ and $T_0$ for $l\geq 2$, even.
\item $r^l$ and $T_1$ for $l\geq 3$, odd.
\end{itemize}

\item In the shells we have to take into account the two usual homogeneous 
solutions, which are not singular in this case. We could remark that if 
$r^l$ is exactly described by the Chebyshev expansion, and so implies an 
exact zero determinant for the matrix, it is not the case for the fractional 
solution $r^{-\left(l+1\right)}$. This one is not given by a finite sum of Chebyshev 
polynomials but rather by an infinite sum implying the result would be worse 
and worse as the number of coefficients 
increases, for the determinant of the matrix would be closer and closer to $0$.
So, to deal with this, we have to take into account that the eigenvalue $0$ is 
of order $2$ even if $r^{-\left(l+1\right)}$ only  becomes an exact eigenvector
for an infinite number of coefficients.

\item In the external domain, the solution $r^l$ is singular at infinity 
except for $l=0$. $r^{-\left(l+1\right)}$ is always acceptable.

If the source is multiplied by $r^4$, the finite part introduce two others 
eigenvectors of eigenvalue $0$~: $T_0$ and $T_1$. So the situation is~:
\begin{itemize}
\item $T_0$ and $T_1$ for $l=0$.
\item $T_0$, $T_1$ and $r^{-\left(l+1\right)}$ for $l\geq 1$.
\end{itemize}

If the source is multiplied by $r^3$, the finite part only introduce one other
eigenvector of eigenvalue $0$~: $T_0$, and the situation is~:
\begin{itemize}
\item $T_0$ and $r^{-\left(l+1\right)}$ for all $l$.
\end{itemize}

If the source is multiplied by $r^2$, they are no other solutions than the usual 
ones which gives~:
\begin{itemize}
\item $T_0$ and $T_1$ for $l=0$.
\item $r^{-\left(l+1\right)}$ for $l\geq 1$.
\end{itemize}
\end{itemize}

From the above discussion we are able to determine the order $p$ of the eigenvalue $0$.
The banded-matrices are then amputated from their $p$ first columns and their
 $p$ last lines resulting in invertible banded-matrices. We abandon the $p$
last coefficients of the source. Doing so, we find a particular solution 
of the system which has its $p$ first coefficients undefined and thereafter 
set to zero.

\noindent In particular the previously presented matrix (in the kernel, 
for $l=2$ and $N=9$) becomes~:
$$
\pmatrix{
56&-336&-200&0&0&0&0\cr
56&96&-488&336&0&0&0\cr
0&144&168&-672&-504&0&0\cr
0&0&264&272&-888&-704&0\cr
0&0&0&416&408&-1136&-2000\cr
0&0&0&0&600&1392&1488\cr
0&0&0&0&0&816&1840\cr}
$$

Before solving the system, an $LU$ decomposition is performed using LAPACK {\it (Linear 
Algebra PACKage)} \cite{Lapack} for purpose of rapidity. LAPACK is also used for the 
resolution of the system.
 
\subsection{Regularity and boundary conditions}
In this section we will show how some homogeneous solutions are used to 
maintain regularity and satisfy the boundary conditions. We will concentrate on
 the boundary condition $f=0$ at infinity.

\begin{itemize}
\item In the kernel, the operator is singular only for $l \geq 2$. If it is the
case, to maintain regularity, $f$ has to verify the 
following conditions~:
\begin{eqnarray}
f\left(0\right) &=& 0 \\
f'\left(0\right) &=& 0 
\end{eqnarray}
Thanks to the parity of the Chebyshev expansion, one of these conditions is 
always fulfilled, depending on the parity of $l$. So we perform a linear 
combination of the solution with either $T_0$ or $T_1$ to fulfill the 
other one. Nothing has to be done for $l\leq 1$.

\item In the shells, nothing has to be done for there are neither boundary 
conditions nor singularities.

\item In the external domain we should, once more, discriminate between three 
cases :

\begin{itemize}
\item If the source is multiplied by $r^4$, we must impose $f\left(1\right) = 0$ 
to satisfy the boundary condition ; this is done by performing a linear 
combination of the solution and $T_0$. Then for $l\geq 1$, for reasons of regularity, 
we must have $f'\left(1\right) = 0$, a condition which is obtained by linear
 combination with $T_1$.

\item In the case of a source multiplied by $r^3$, the boundary condition 
$f\left(1\right) = 0$ ensures regularity ; we impose it by performing a linear 
combination of the solution and $T_0$.

\item If the source is multiplied by $r^2$, the situation is a bit more subtle. 
There is no condition of regularity, but the boundary condition imposes that 
$f\left(1\right)=0$. Then one can show that this implies that 
the source decreases as $r^{-3}$ at infinity.
Conversely, if the source decreases as $r^{-3}$, it 
implies, for $l \not= 0$, that $f\left(1\right)=0$.

So to verify boundary conditions we only consider sources decreasing as 
$r^{-3}$. It implies that the boundary condition is 
automatically verified for $l \not= 0$. We only impose it for $l=0$, by 
doing a linear combination of the solution and $T_0$.

Let us emphasize that this is only the theoretical aspect of the problem. 
During an actual physical calculation, the sources of the Poisson equation 
are themselves numerically given so that they might, due to computational 
errors, not decrease exactly like $r^{-3}$. In such a case one should be 
cautious, for the solution will not exactly be zero at infinity. A possible 
treatment is to enforce the $r^{-3}$ decay by slightly modifying the source 
prior to the resolution of the Poisson equation.
\end{itemize}
\end{itemize}

\subsection{Continuity}
At this stage, for each $\left(l,m\right)$, we are left with a particular 
solution in each domain, one homogeneous solution in the kernel and in the 
external domain, and two in each shell. The last linear combinations will 
be performed to ensure the continuity of the solution and of its first derivative 
across each boundary.

The simplest case is when the angular sampling is the same in every domain (i.e. the 
same numbers of point in $\theta$ and $\phi$). The unknowns are the 
coefficients of the homogeneous solutions in the physical solution and the 
equations are given by matching $f$ and its derivative across each boundary.
 It is easy to see that there is exactly the same number of equations and of 
unknown quantities, resulting in a uniquely determined solution.

If the angular sampling is not the same, the situation is a bit more complex, for 
some $Y_l^m$ may not be present in some domains. At each boundary, 
for each $\left(l,m\right)$, three situations can occur~:
\begin{itemize}
\item the harmonic is present in the two domains : we perform the matching 
of both $f$ and its derivative.
\item the harmonic is present in neither domains : no equation is written.
\item the harmonic is present only in one domain : we assure the continuity 
of $f$ supposing that the harmonic has its coefficient equal to $0$ in the 
domain where it is not present. We perform no matching for its derivative.
\end{itemize}
This procedure results in a system of equations that admit a unique set of 
solutions. We have imposed exactly as much continuity as the sampling 
allowed us.

\noindent To illustrate this, let us take the situation given by Tab. 
\ref{seule_table}, for a specific value $\left(l,m\right)$~:

\begin{table}[ht]
\label{seule_table}
\caption{Example of the situation before doing the connection across 
each boundary}
\begin{tabular*}{\textwidth}{@{\extracolsep{\fill}}cccccc}
\hline
Domain & Bounds & $Y_l^m$ & Particular&
Homogeneous& Unknowns \cr
 & & & solutions & solutions \cr
\hline
$0$ & $0\leq r\leq R_1$ & Yes & $f_0$ & $r^l$ & $\alpha_0$ \cr
$1$ & $R_1\leq r\leq R_2$&Yes & $f_1$ & $r^l$ and $r^{-\left(l+1\right)}$
 & $\alpha_1$ and $\beta_1$ \cr
$2$ & $R_2\leq r\leq R_3$&No \cr
$3$ & $R_3\leq r\leq R_4$&No \cr
$4$ & $R_4\leq r\leq \infty$ & Yes & $f_4$ & $r^{-\left(l+1\right)}$ & 
$\beta_4$ \cr
\hline
\end{tabular*}
\end{table}

In that situation the domain $0$ is the kernel and the domain $4$ is the external 
compactified region. The column labeled $Y_l^m$ denotes the presence or the absence 
of the considered spherical harmonic in each domain. The particular and homogeneous 
solutions are expressed taking into account the sampling and the nature of each 
domain. The unknowns are the coefficients of the homogeneous solutions labeled 
$\alpha$ for $r^l$ and $\beta$ for $r^{-\left(l+1\right)}$. Using the procedure 
described above we obtain the following equations~:

\begin{itemize}
\item For $r = R_1$, the spherical harmonic is present in both domains, so 
we have to write the continuity of the solution and its derivative, which gives
\begin{eqnarray}
f_0\left(R_1\right) + \alpha_0 R_1^l &=& f_1\left(R_1\right) + \alpha_1 R_1^l
+ \beta_1 R_1^{-\left(l+1\right)} \\
f'_0\left(R_1\right) + l \alpha_0 R_1^{\left(l-1\right)} &=& 
f'_1\left(R_1\right) + l \alpha_1 R_1^{\left(l-1\right)} - \left(l+1\right)
\beta_1 R_1^{-\left(l+2\right)}.
\end{eqnarray}

\item For $r=R_2$, the spherical harmonic is present only in the domain $1$ and so 
we write only the continuity of the solution assuming that it is zero in the domain 
$2$
\begin{equation}
f_1\left(R_2\right) + \alpha_1 R_2^l + \beta_1 R_2^{-\left(l+1\right)} = 0.
\end{equation}

\item For $r=R_3$, no equation is written, for the harmonic is absent on both sides 
of the boundary.

\item For $r=R_4$, the situation is the same as at $r=R_2$

\begin{equation}
f_4\left(R_4\right) + \beta_4 R_4^{-\left(l+1\right)} = 0.
\end{equation}
\end{itemize}

We have now four independent equations which are solved to find the unknowns 
$\alpha_0$, $\alpha_1$, $\beta_1$ and $\beta_4$.

All this procedure enables us to find a unique solution of the scalar 
Poisson equation, solution to be regular everywhere, continuous, like its 
derivative, and that is zero at infinity. We should point, once more, 
that the source must decrease at least as $r^{-3}$, for this to be possible.

\section {Convergence properties of the scalar Poisson equation solver}
\label{secconv}
\subsection{Position of the problem}
We wish to study the convergence of our algorithm, depending on the number of 
coefficients chosen for the $r$-expansion. The number of points for $\theta$ and
$\phi$ does not change the precision of the result, as long as we have enough 
points, that is enough spherical harmonics, to describe the source 
properly.
However, concerning $r$, we perform matrix-inversion and we expect a better 
precision as the number of coefficients increases.

It is well known (see \cite{CanutHQZ88, GottlO77}) that with spectral 
method, the error is evanescent, i.e. decreasing 
as $\exp\left(-N\right)$, $N$ being the number of coefficients, as long as 
we are working with functions that are $\cont{\infty}$. If the functions are 
only $\cont{p}$, the error is decreasing as $N^{-\left(p+1\right)}$ solely. 
This is known as the Gibbs phenomenon.

The various domains of our multi-domains method \cite{BonazGM98}
are intended to fit 
the surfaces of discontinuity, for example the surface of a star (see 
\cite{GourgHLPBM99} for an application to stars with discontinuous 
density profiles, like strange stars). Doing so, 
each function is $\cont{\infty}$ in each domain, removing any Gibbs 
phenomenon.

To test the validity of our numerical scheme, we compare calculated solutions to 
analytical ones. We estimate the relative error as the infinite norm of the 
difference over the infinite norm of the analytical solution. We will present 
some examples for the construction of analytical solutions. Using only 
$\cont{\infty}$ functions we expect errors to be evanescent.

But this is not so simple. It can be easily shown that, generally, the particular 
solutions obtained by the inversion of the operator for each 
$\left(l,m\right)$ are of polynomial or fractional type. Those functions 
are exactly described by Chebyshev polynomials in 
$r$ or $r^{-1}$. This is true except in two cases, related to the 
homogeneous solutions~:

\begin{itemize}
\item a source in $r^{l-2}$, will give rise to a particular solution 
in $r^l \ln r$.
\item a source in $r^{-\left(l+3\right)}$, will be associated with a 
particular solution in $r^{-\left(l+1\right)}\ln r$.
\end{itemize}

In such cases, we expect some problems, for the description of logarithm 
functions in terms of Chebyshev polynomials may not be accurate. To be 
more precise about this effect, let us study the situation in each type 
of domain.

\subsubsection{In the kernel}

In the kernel and for reason of regularity, sources in 
$r^{-\left(l+3\right)}$ are obviously never present. At first sight, the 
case of a source in $r^{l-2}$ seems to be more problematic~;~but let us 
recall that this source has to be the factor of $Y_l^m$. Can we have 
a source containing terms like $Y_l^m r^{l-2}$~? To answer this question 
we refer to \cite{BonazGM99b} where it is shown that, for a regular 
function (i.e. function expandable as a polynomial series in Cartesian 
coordinates $\left(x,y,z\right)$ associated with $\left(r,\theta,\phi\right)$),
 terms like $r^\alpha Y_l^m$ are present in the spectral expansion 
only if $\alpha \geq l$. So, sources leading to a $\ln$ function in the kernel
are not regular at the origin. To conclude, we expect no problem connected 
with particular solutions containing $\ln$ functions in the kernel, at least 
with physical regular sources. However, let us mention the fact that if the source 
is the result of some calculation, it might contain some unphysical terms due to 
computational errors. Those terms might give rise to some logarithmic functions.
\subsubsection{In the shells}

As usual there are no regularity prescriptions in the shells. The two 
types of particular solutions can appear. 
To investigate more precisely the effects of the logarithm, we studied the 
behavior of the error performed by expanding the two types of particular 
solutions in Chebyshev polynomials.

We constructed the two following exact particular solution $r^l \ln r$ 
and $r^{-\left(l+1\right)}\ln r$, approached them by a sum of Chebyshev 
polynomials in $x$, the relation between $r$ and $x$ being $r= \alpha x 
+ \beta$. Then, we estimate the error by the same method as the one 
described before.

\begin{figure}
\centerline{\includegraphics[height=9cm,angle=-90]{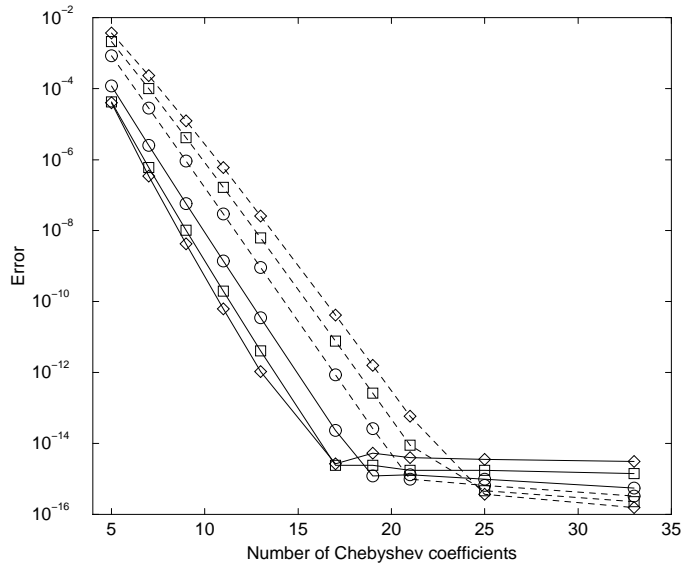}}
\vskip12pt
\caption{\label{erreur_log_shell}
Relative difference (infinite norm) between the particular solutions with 
logarithm and their truncated Chebyshev expansions, in a shell.

The scale for the number of coefficients is linear.

The solid lines represent the $r^l \ln r$ functions and the dashed lines 
the $r^{-\left(l+1\right)}\ln r$ ones.

The circles represent the case $l=0$, the squares $l=1$ and the diamonds 
$l=2$.

This plot has been obtained using $\alpha = 0.5$ and $\beta = 1.5$.
} 
\end{figure}

Fig. \ref{erreur_log_shell} shows an evanescent error. The functions 
containing logarithm are thus rather well described in a shell. This is due 
to the fact that the $\ln$ functions are bounded in such domains and not 
going to infinite values. More precisely, the $r^l \ln r$ and $r^{-\left(l+1\right)}
\ln r$ functions are $\cont{\infty}$ in the shells so that the error should 
be evanescent. Let us mention that this result does not 
depend on the choice made for $\alpha$ and $\beta$. To conclude we 
expect no problem to rise from the presence of such particular solutions 
in the shells.

\subsubsection {In the external compactified domain}\label{origine}

The particular solution in $r^l \ln r$ is not going to zero at 
infinity, and so cannot appear in the external domain. But the other type 
of particular solutions $r^{-\left(l+1\right)} \ln r$ is likely to appear. 
We investigate their effect by the same method than the one used in the 
shells, that is determining the behavior of the error done by 
interpolating the exact solution by a finite sum of Chebyshev polynomials.

\begin{figure}
\centerline{\includegraphics[height=9cm,angle=-90]{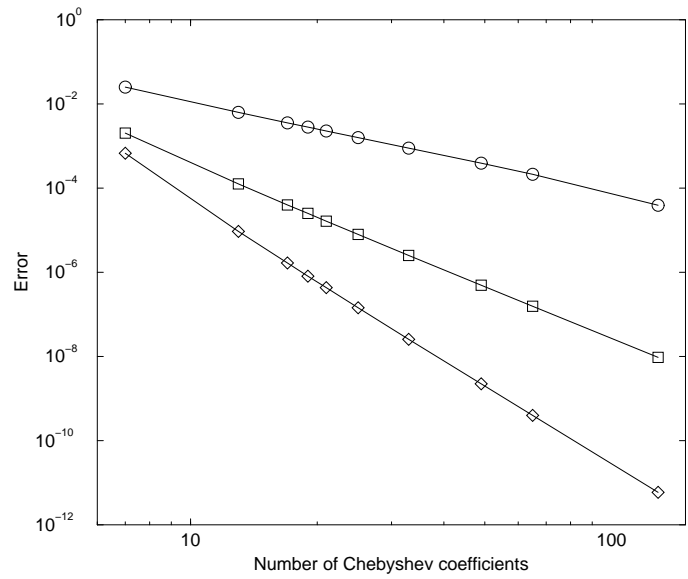}}
\vskip12pt
\caption{\label{erreur_log_zec}
Relative difference (infinite norm) between the particular solutions with 
logarithm and their truncated Chebyshev expansions, in the external domain.

The scale for the number of coefficients is logarithmic.

The circles represent the case $l=0$, the squares $l=1$ and the diamonds 
$l=2$.
} 
\end{figure}

\begin{figure}
\centerline{\includegraphics[height=9cm,angle=-90]{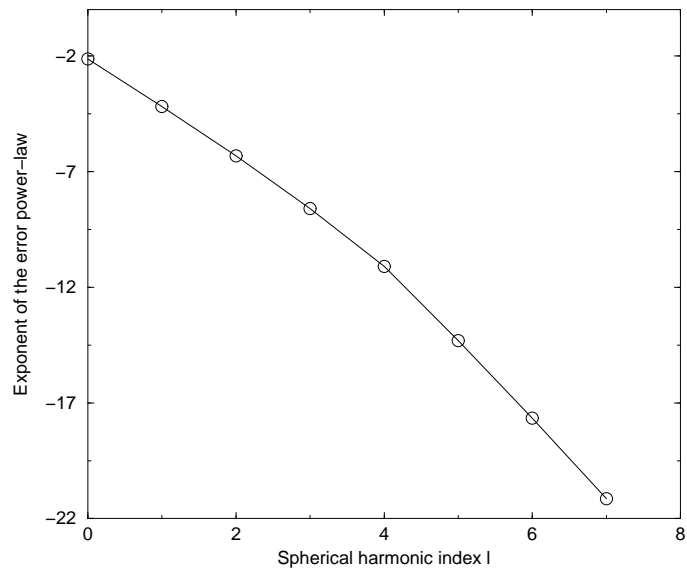}}
\vskip12pt
\caption{\label{pentelog}
Exponent of the power-law followed by the error shown in Fig. 
\ref{erreur_log_zec}, as a function of $l$.}
\end{figure}

Fig. \ref{erreur_log_zec} shows that the error is no longer evanescent but
follows a power law. The error is decreasing faster and faster 
as $l$ increases, for the associated particular solution is being 
better approached by Chebyshev polynomials. In other words, the function 
$r^{-\left(l+1\right)}\ln r = -u^{\left(l+1\right)}\ln u$ is not $\cont{\infty}$, 
for its $\left(l+1\right)^{{\mathrm th}}$ derivative contains terms in $\frac{1}{u}$, not 
regular at spatial infinity, that is for $u=0$.
More precisely, Fig. \ref{pentelog} shows the value of the exponent as a 
function of $l$.

We can conclude that the error done by expanding $r^{-\left(l+1\right)} 
\ln r$ in Chebyshev polynomials follows a power-law and that it is 
decreasing faster than $N^{-2\left(l+1\right)}$. 
We will use this to explain some features of our scalar and vectorial 
Poisson equation solvers.

\subsection{Accuracy estimated by comparison with analytical solutions}
From the results of the previous section, we expect an evanescent error 
for the resolution of the scalar Poisson equation when there is no 
particular solution containing any logarithm in the external domain and an 
error following a power-law when such solutions appear. We present here 
some results that illustrate this behavior and lead to 
two properties about the error.

\subsubsection{Spherically symmetric source}
First of all, let us consider a simple case for which we do not expect any 
Gibbs-like phenomenon~: a spherically symmetric source decreasing as $r^{-4}$. In fact, 
the only harmonic present in this source is $l=0$, that would imply a 
$\ln$ solution only for a source in $r^{-3}$. We choose a source $S$ decreasing 
as $r^{-4}$ in the external domain and a polynomial one, such that the solution 
is not singular in the kernel. The associated solution $F$ can be found analytically.

\noindent In the external domain, for $r>R$, we have
\begin{equation}
S = \frac{R^5}{r^4} \phantom{mm} {\mathrm ;} \phantom{mm}
F = \frac{R^5}{2r^2}-\frac{17}{15}\frac{R^4}{r},
\end{equation}
and for $r<R$,
\begin{equation}
S = R-\frac{r^2}{R} \phantom{mm} {\mathrm ;} \phantom{mm}
F = \frac{Rr^2}{6}-\frac{r^4}{20R}-\frac{3}{4}R^3.
\end{equation}

\begin{figure}
\centerline{\includegraphics[height=9cm,angle=-90]{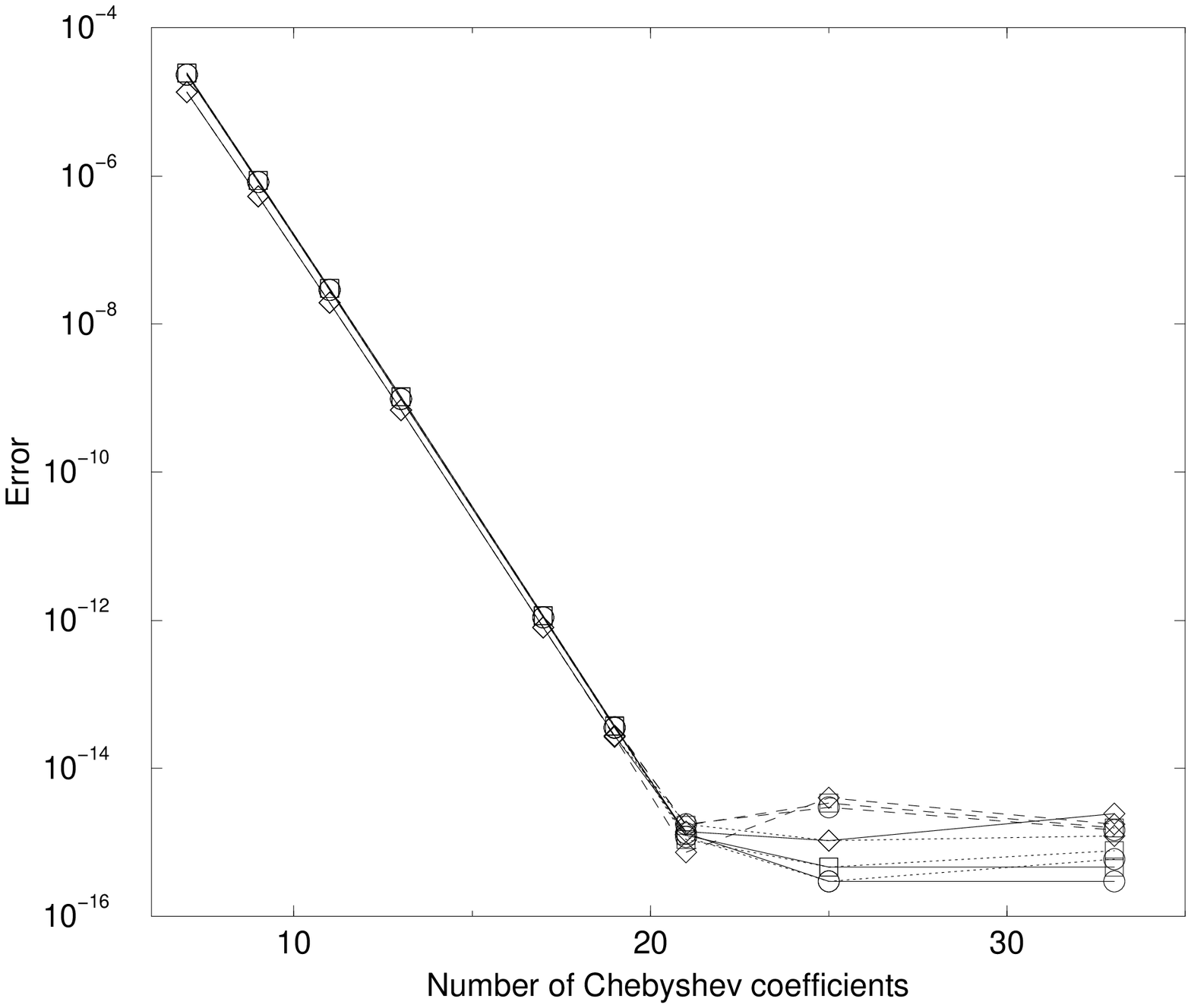}}
\vskip12pt
\caption{\label{sphere} Error on the resolution of the scalar Poisson equation for a 
spherically symmetric source extending to infinity.

The solid lines represent the $r^4 S$ scheme, the dotted lines the $r^3 S$ scheme 
and the dashed lines the $r^2 S$ one.

The scale for the number of coefficients is linear.

The circles represent the error in the kernel, the squares in the shell and 
the diamonds in the external domain.}
\end{figure}

As expected Fig. \ref{sphere} shows an evanescent error, with some 
saturation at the level of $10^{-15}$ due to the round-off error, the calculation 
being performed in double precision. No significant difference can be seen 
between the three schemes.

\subsubsection{Compact source}\label{compact}
Another interesting case is that of a source with a compact support, that is a 
source which is zero in the external domain. As for the previous case we do not 
expect any Gibbs phenomenon. In the external domain let us choose the 
following analytical solution
\begin{equation}
F = Y_l^0 \frac{1}{r^{l+1}}.
\end{equation}
This solution leads to a source that vanishes in the external domain. To 
avoid any singularity at the center, we choose the later function as a 
solution of the equation for $r<R$
\begin{equation}
F = Y_l^0 \left[\left(2l+5\right)\frac{r^2}{2R^{2l+3}} -
		\left(2l+3\right)\frac{r^4}{2R^{2l+5}}\right].
\end{equation}

This solution has been chosen so that $F$ and its first derivative with respect
to $r$ are continuous at $r=R$, properties of the solution given by our 
algorithm. The 
associated source, for $r<R$, is found by taking the Laplacian of $F$
\begin{equation}
S = Y_l^0 \left[\left(2l+5\right)\left(2l+3\right)\frac{1}{R^{2l+3}}
	-\left(2l+3\right)\left(4l+10\right)\frac{r^2}{R^{2l+5}}
		\right].
\end{equation}

So we constructed a non-spherically symmetric compact source, which contains only one 
spherical harmonic. We chose for simplicity $m=0$, for we do not expect 
any variation with $m$, the later being absent of the inverted operator.

\begin{figure}
\centerline{\includegraphics[height=9cm,angle=-90]{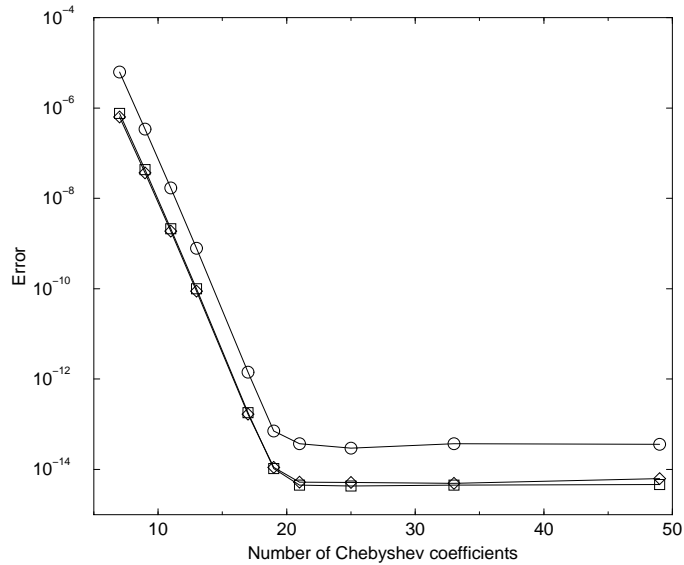}}
\vskip12pt
\caption{\label{comp_2} Error on the resolution of the Poisson like equation for a 
non-spherical compact source with $l=2$.

The scale for the number of coefficients is linear.

The circles represent the error in the kernel, the squares in the shell and 
the diamonds in the external domain.}
\end{figure}

As expected, Fig. \ref{comp_2} shows an evanescent error down to a 
saturation value of approximatively $10^{-14}$.

\subsubsection {A logarithm in a shell}
The last case with an evanescent error we considered is the one where the problematic 
particular solutions (i.e. containing a logarithm) appear only in a shell 
bounded by $R_1<r<R_2$. We choose 
a source $s$ that implies the appearance of both type of particular solutions.
Let $F$ be the associated solution. In the shell, for $R_1<r<R_2$, we have
\begin{eqnarray}
S_{\mathrm shell} &=& \frac{1}{r^3}+3\frac{z^2}{r^2}-1 \\
\nonumber
F_{\mathrm shell} &=& -\frac{\ln r}{r}+\frac{\ln R_1 -1}{r}+\frac{1}{R_2}
+\left(3\frac{z^2}{r^2}-1\right)\left(\frac{1}{5}r^2\ln r - 
\left(\frac{\ln R_2}{5}+\frac{1}{25}\right)r^2 +\frac{R_1^5}{25}\frac{1}{r^3}\right)
.\end{eqnarray}

For simplicity, we take $S=0$ in the kernel and in the external domain, the 
solution being chosen, once more, by continuity across the boundaries
\begin{eqnarray}
F_{\mathrm kernel} &=& \frac{1}{R_2}-\frac{1}{R_1}+\frac{\ln R_1 - \ln R_2}{5}r^2
\left(3\frac{z^2}{r^2}-1\right) \\
\nonumber
F_{\mathrm external} &=& \frac{\ln R_1 -\ln R_2}{r}+\frac{1}{r^3}
\frac{R_1^5-R_2^5}{25}\left(3\frac{z^2}{r^2}-1\right).
\end {eqnarray}

\begin{figure}
\centerline{\includegraphics[height=9cm,angle=-90]{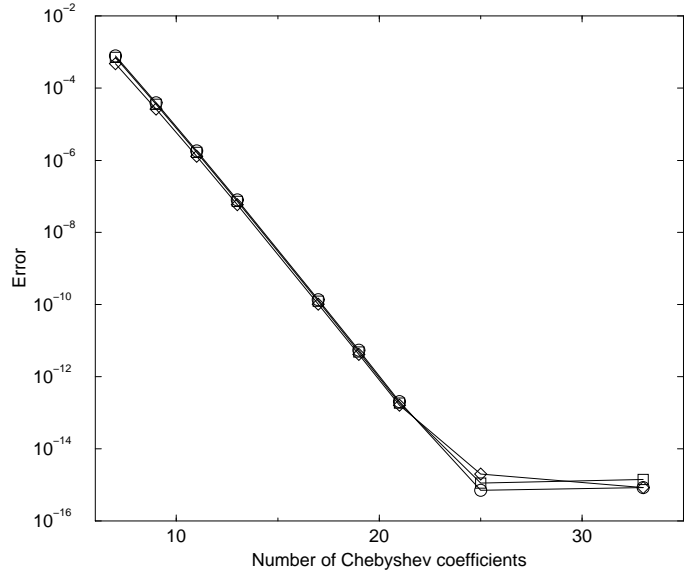}}
\vskip12pt
\caption{\label{log_shell} Error on the resolution of the scalar Poisson equation for a 
solution containing bounded logarithm functions.

The scale for the number of coefficients is linear.

The circles represent the error in the kernel, the squares in the shell and 
the diamonds in the external domain.}
\end{figure}

 The result presented in Fig. \ref{log_shell} shows an evanescent error, confirming that the 
presence of logarithm function is only a problem in the external domain. Once more 
let us mention that this is due to the fact that the logarithm functions 
are bounded in a shell and not going to infinite values.
Such bounded functions are rather well described in terms of Chebyshev 
polynomials.

\subsubsection {The Gibbs phenomenon}
Let us now consider a case where the particular solution contains a logarithm in 
the external domain. 
Following the construction of the source and solution of Sec. \ref{compact}
let us take the following source in the external domain
\begin{equation}
S = -Y_l^0 \frac{1}{r^{l+3}},
\end{equation}
and $S = 0$ for $r<R$.
The associated unique solution is
\begin{eqnarray}
F &=& Y_l^0 \frac{r^l}{\left(2l+1\right)^2 R^{2l+1}} \phantom {ppppp}
{\mathrm for} \phantom{m}r<R \\
\nonumber
F &=& Y_l^0 \frac{\ln\left(r\right)-\ln\left(R\right)+\frac{1}{2l+1}}
		{\left(2l+1\right)r^{l+1}}\phantom {ppppp}
{\mathrm for} \phantom{m}r>R.
\end{eqnarray}

\begin{figure}
\centerline{\includegraphics[height=9cm,angle=-90]{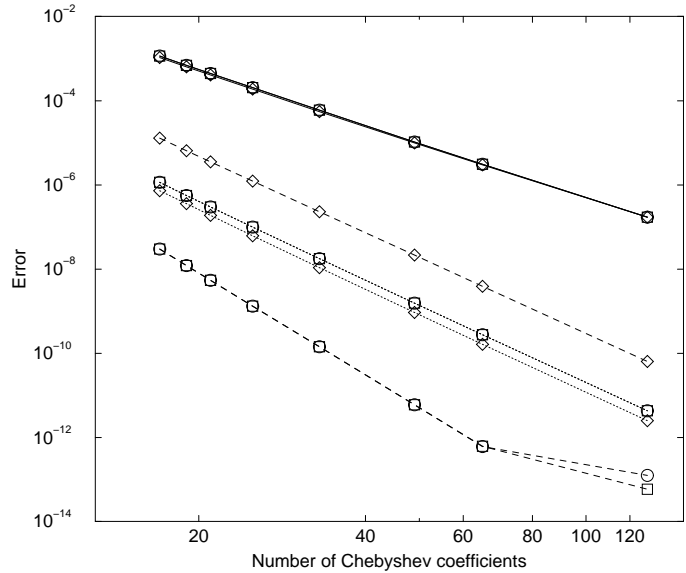}}
\vskip12pt
\caption{\label{gibbs_2_4_2} Error on the resolution of the scalar Poisson equation for a 
solution containing $\ln$ functions for $l = 2$.

The scale for the number of coefficients is logarithmic.

The circles represent the error in the kernel, the squares in the shell and 
the diamonds in the external domain.

Solid lines represent the scheme with $r^4 S$, the dotted ones the scheme 
with $r^3 S$ and dashed lines the one with $r^2S$.}
\end{figure}
Fig. \ref{gibbs_2_4_2} presents an example of the obtained result for each 
of the three schemes discussed in Sec. \ref{matrix}.
A logarithm being present in the solution, the error is no longer evanescent and 
follows a power-law. One important feature is that the $r^4S$ scheme
is converging much less rapidly than the $r^3S$ and $r^2S$ ones. It may come from
 the fact that for a given source, the $r^4 S$ scheme is dealing with particular solutions
less rapidly decreasing.

\begin{figure}
\centerline{\includegraphics[height=9cm,angle=-90]{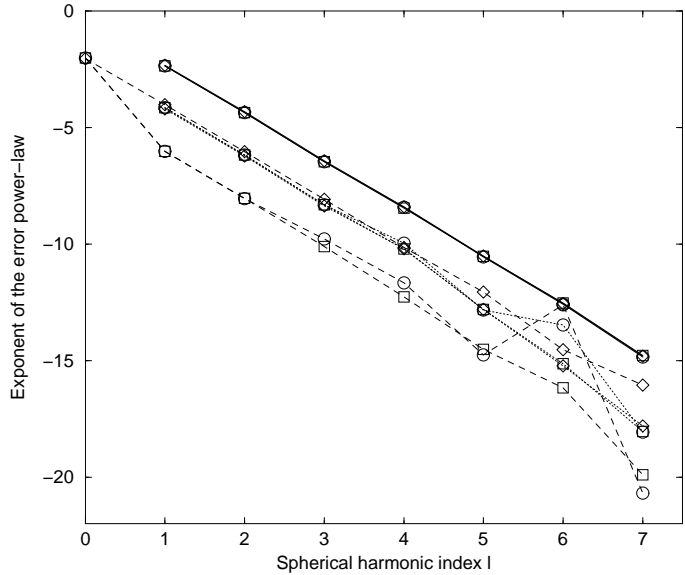}}
\vskip12pt
\caption{\label{pente} 
Exponent of the power-law followed by the error shown in Fig. \ref{gibbs_2_4_2}
as a function of the index $l$.

The solid lines correspond to the $r^4 S$ scheme, the dotted lines to the $r^3S$ one 
and the dashed lines to the $r^2 S$ one.

The circles represent the error in the kernel, the squares in the shell and 
the diamonds in the external domain.}
\end{figure}

In Fig. \ref{pente} the slope of the 
power-law is plotted as a function of the harmonic index $l$, for the three different 
schemes. It reveals an error decreasing as $N^{-2\left(l+1\right)}$ for the 
$r^2S$ and $r^3S$ schemes and as $N^{-2l}$ for the $r^4S$ one. Let us mention that the 
$r^2S$ scheme yields an error following the same power-law than the one 
rising from the description of the associated function (cf. Sec. \ref{origine})
, making us confident about the origin of such a behavior.

\subsection{Convergence properties} \label{converge}
All the examples shown in the previous section enable us to propose the 
two following empirical properties concerning the decrease of the error.

\begin{itemize}
\item{\bf Property 1 : }if the source is decreasing as $r^{-k}$ 
at infinity and does not contain any spherical harmonics with
 $l\geq k-3$, then the error is evanescent.

\item{\bf Property 2 : }if the source decreases at least as $r^{-k}$ 
at infinity, then the error decreases at least as
$N^{-2\left(k-2\right)}$ (resp. $N^{-2k}$) for the $r^2S$ and $r^3S$ schemes 
(resp $r^4S$ scheme).
\end{itemize}

The first one is just issued from the condition to have a $\ln$ function 
in the external domain and the second one from the values of the power-law 
found in the previous section.

\pagebreak
\section{Vectorial Poisson equation}\label{secvect}

Using the Poisson equation solver from Sec. \ref{scalar_eq} and studied in Sec. 
\ref{secconv}, we focus now on the vectorial 
Poisson equation given by Eq. (\ref{e:poisson_vect}), in the non-degenerated 
case (i.e. $\lambda \not= -1$).

Let us first mention that the operator 
$\Delta + \lambda \vec{\nabla}(\vec{\nabla}\cdot \ )$
has been shown to be strongly elliptic and self-adjoint in 
Refs.~\cite{York73,York74}
in the case $\lambda=1/3$ (conformal Laplace operator).
Conditions for existence and uniqueness of
solutions have been presented in Appendix B of Ref.~\cite{SmarrY78}. The
harmonic vectorial functions of this operator and the associated
multipole expansions have been discussed by \'O Murchadha
\cite{Omurc92}.

Three different schemes have been previously proposed by 
other authors \cite{BowenY80, OoharN97, OoharNS97} to
reduce the resolution of Eq.~(\ref{e:poisson_vect}) to those of four 
scalar Poisson equations. Let us emphasize that those three schemes are not 
covariant. They are only applicable in Cartesian coordinates which allow us to 
commute operators like Laplacian and gradient.

Let us mention the fact that a different method, based on solving for 
the degenerated case (i.e. $\lambda = 1$) has been proposed in 
\cite{BonazGM99b} but is not studied in the present work.

\subsection {The Bowen-York method}

The idea of this method (see \cite{BowenY80}) is to search for the 
solution of Eq. (\ref{e:poisson_vect}) in the form
\begin{equation}
\vec{N} = \vec{W} + \vec{\nabla} \chi,
\end{equation}
where $\vec{W}$ and $\chi$ are solutions of
\begin{eqnarray}
\Delta \vec{W} &=& \vec{S} \\
\Delta \chi &=& -\frac{\lambda}{\lambda+1}\vec{\nabla}\cdot \vec{W}.
\end{eqnarray}

This method gives a solution to Eq. (\ref{e:poisson_vect}) but let us check that this
solution is the one that is $\cont{1}$. $\vec{W}$ is $\cont{1}$, being 
solution of a Poisson equation. This implies that the source of the equation 
for $\chi$ is continuous, and that $\chi$ is $\cont{2}$. This is sufficient 
to ensure that $\vec{N}$ is $\cont{1}$. The scheme finds the only solution 
$\cont{1}$ and going to zero at infinity.

Unfortunately this very simple method is not applicable with our 
Poisson equation solver, for the physical sources are not decreasing fast 
enough at 
infinity. For the problem that motivated this study,
namely binary neutron star systems \cite{BonazGM99a,GourgGTMB01}, 
the source $\vec{S}$ of Eq. (\ref{e:poisson_vect}), is expected to behave like
$r^{-4}$ at infinity implying that we can calculate $\vec{W}$. This vector 
field is acting like $r^{-1}$ at infinity, for 
$r^{-1}$ is an homogeneous solution of the scalar Poisson 
equation usually present (monopolar term).

So the source of the equation for $\chi$, being the divergence of $\vec{W}$, behaves
like $r^{-2}$. This decreasing is not fast enough to compute 
the value of $\chi$. Analytically no problem occurs because only the gradient of
$\chi$ is relevant, not $\chi$ itself, for the calculation of the solution. 
To summarize, the implementation of this scheme conducts to the computation of 
diverging quantities, making the result wrong in the external domain. We should say 
that this scheme is applicable for domains not extending to infinity. 
However it may be possible to use it by treating analytically the 
diverging quantities.

\subsection{The Oohara-Nakamura method}
In this case (see Sec. $3.1.1$ of \cite{OoharN97}) we start by solving the following scalar equation
\begin{equation}
\label{chioohara}
\Delta \chi = \frac{1}{\lambda + 1} \vec{\nabla} \cdot\vec{S}.
\end{equation}
Then the solution of Eq. (\ref{e:poisson_vect}) is found by solving the 
following set of three equations
\begin{equation}
\label{eqoohara}
\Delta \vec{N} = \vec{S} - \lambda \vec{\nabla}\chi.
\end{equation}

Comparing (\ref{e:poisson_vect}) with (\ref{eqoohara}) shows that this scheme 
gives the exact solution of Eq. (\ref{e:poisson_vect}) if and only if
\begin{equation}
\label {conditionoohara}
\vec{\nabla} \chi = \vec{\nabla}\left(\vec{\nabla}\cdot\vec{N}\right).
\end{equation}
But the scalar equation (\ref{chioohara}) only ensures that
\begin{equation}
\label{lapoohara}
\Delta\left(\chi-\vec{\nabla}\cdot\vec{N}\right) = 0.
\end{equation}

From the study of Sec. \ref{scalar_eq}, we can show that it is possible to 
construct an homogeneous solution of the scalar Poisson equation, in all 
space, that is non-zero, going to zero at infinity, if and only if, that 
solution is not $\cont{1}$.

In the general case, $\vec{\nabla}\cdot\vec{N}$ is only $\cont{0}$ at boundary between the 
different domains, while $\chi$, solution of a Poisson equation, is 
$\cont{1}$. So it is possible to fulfill Eq. (\ref{lapoohara}) and not 
Eq. (\ref{conditionoohara}). If $\vec{\nabla}\cdot\vec{N}$ is $\cont{1}$, then Eq. 
(\ref{lapoohara}) implies, as shown by Sec. \ref{scalar_eq}, that~:
$\chi = \vec{\nabla}\cdot\vec{N}$. In this case, the condition 
(\ref{conditionoohara}) is trivially fulfilled. Imposing that 
$\vec{\nabla}\cdot\vec{N}$ is at least $\cont{1}$, is equivalent to impose 
that $\vec{S}$ is continuous across every boundary.

To conclude, let us say that the Oohara-Nakamura method gives the exact solution 
if and only if the source $\vec{S}$ is continuous across every boundary 
delimiting the different domains. This property is general, meaning that it is 
not due to our numerical method. We can mention that the found solution 
is the $\cont{1}$ one, because it is calculated as solution of three 
scalar Poisson equations.

Second let us see if this scheme is applicable, using our scalar Poisson 
equation solver. At first sight, this scheme suffers the same drawback than the Bowen-York one.
Because of homogeneous solutions of scalar Poisson equation, $\chi$ is 
decreasing as $r^{-1}$ at infinity and its gradient as $r^{-2}$, which is 
not enough to allow us to solve the set (\ref{eqoohara}) of three scalar 
Poisson equations.

The difference is that the solution of Eq. (\ref{eqoohara}) is the solution 
of the vectorial Poisson equation (\ref{e:poisson_vect}) and we must be able to set it 
to zero at infinity, contrary to the Bowen-York method where the problem 
occurs for auxiliary quantities.

So it must be possible to show that the source of Eq. (\ref{eqoohara})
decreases fast enough, that is, at least as $r^{-3}$. The problem arises 
from the monopolar term of $\chi$, i.e. the only one that gives an 
homogeneous solution in $r^{-1}$ in the external domain.
It is known, that the mono-polar term $M_0$ of the solution of a scalar 
Poisson equation with source $\sigma$, is given by
\begin{equation}
M_0 = \frac{1}{4 \pi} \int\!\!\!\int\!\!\!\int \sigma {\mathrm d}^3r,
\end{equation}
the integration being performed over all space.

\noindent Now we have $\sigma = \vec{\nabla}\cdot \vec{S}$. The use of 
Green formula leads to
\begin{equation}
M_0 = \frac{1}{4 \pi} \int\!\!\!\int \vec{S} \cdot {\mathrm d}\vec{s},
\end{equation}
the surface integration being done at infinity. But $\vec{S}$ decreases as 
$r^{-4}$, implying that the surface integral is zero, that is $M_0 = 0$.
This remains true if the source acts only like $r^{-3}$.

So the monopolar term of $\chi$ is zero, which implies that $\chi$ decreases 
at least as $r^{-2}$. This behavior ensures that the source of Eq. 
(\ref{eqoohara}) decreases as $r^{-3}$, allowing us to find the unique solution
going to zero at infinity.

We implemented and tested this scheme, recalling the reader that it is only 
applicable if the source of Eq. (\ref{e:poisson_vect}) is continuous and 
requires that the source decreases at least like $r^{-3}$ at infinity.

\subsection{The Shibata method}
The solution is now searched as (see \cite{OoharNS97})
\begin{equation}
\label {shiba}
\vec{N} = \frac{1}{2} \frac{\lambda+2}{\lambda+1} \vec{W}
	-\frac{1}{2}\frac{\lambda}{\lambda+1}
	\left(\vec{\nabla} \chi + \vec{\nabla}\vec{W}\cdot \vec{r}\right)
,
\end{equation}
where $\vec{W}$ and $\chi$ are solutions of
\begin{eqnarray}
\Delta \vec{W} &=& \vec{S} \\
\Delta \chi &=& - \vec{r}\cdot\vec{S}.
\end{eqnarray}
and $\vec{r}$ denotes the vector of coordinates $\left(x,y,z\right)$.

This scheme gives a solution to Eq. (\ref{e:poisson_vect}), but, as for the 
Bowen-York
method, let us quickly check that it is the unique $\cont{1}$ going to zero 
at infinity. At infinity, $\vec{W}$, solution of scalar Poisson equation, is 
behaving at least like $r^{-1}$. This ensures that 
$ \vec{\nabla}\vec{W}\cdot \vec{r}$ is zero 
at infinity, proving that the solution goes to zero.

Concerning the continuity, being solutions of scalar Poisson equations, we know 
that both $\vec{W}$ and $\chi$ are at least $\cont{1}$. But we have to take 
care about the term $\vec{\nabla} \chi +\vec{\nabla}\vec{W}\cdot 
\vec{r} $ of Eq. (\ref{shiba}).
First we can show that
\begin{equation}
\Delta \left(\vec{r}\cdot \vec{W}\right) = \vec{r}\cdot\vec{S} + 2
\vec{\nabla}\cdot\vec{W}.
\end{equation}
Using that property and the equation for $\chi$ we can see that
\begin{equation}
\Delta\left(\vec{r}\cdot\vec{W} + \chi\right) = 2\vec{\nabla}\cdot \vec{W}.
\end{equation}
The source of that equation is $\cont{0}$, so that $\vec{r}\cdot\vec{W} + \chi$ is 
$\cont{2}$. The term of Eq. (\ref{shiba}), can be expressed as
\begin{equation}
\label{termapb}
\vec{\nabla} \chi + \vec{\nabla} \vec{W}\cdot \vec{r} = 
\vec{\nabla}\left(\vec{r}\cdot\vec{W}+\chi\right) - \vec{W}.
\end{equation}
Using the continuity properties found above, it is easy to see that the 
right-hand side 
of Eq. (\ref{termapb}) is $\cont{1}$, which ends our demonstration by 
proving that the calculated $\vec{N}$ is $\cont{1}$.

As before, let us now check if this method is applicable by means of our 
scalar Poisson equation solver.
The source of the equation for $\chi$ decreases at least like $r^{-3}$ at 
infinity if and only if $\vec{S}$ decreases like $r^{-4}$. Like the 
Oohara-Nakamura scheme, this one does not involve any diverging 
quantities and so is suitable for numerical purposes.

This method has been implemented and, contrary to the Oohara-Nakamura method, can 
be used even with discontinuous source, but requires that $\vec{S}$ decreases 
at least like $r^{-4}$ at infinity, which, let us recall, is the case for 
the physical problems we intend to study. 

\subsection{Convergence criterion}
As seen before the resolution of Eq. (\ref{e:poisson_vect}) reduces to that of 
four scalar Poisson equations. So we should be able to use the results of Sec.
\ref{converge} to established convergence criterion for the schemes 
proposed in \cite{OoharN97, OoharNS97}.

\subsubsection{The Oohara-Nakamura scheme}
 Let us suppose that the source $\vec{S}$ 
of Eq.~(\ref{e:poisson_vect}) contains only one spherical harmonic $Y_l^m$ and 
decreases as $r^{-k}$ at infinity ($ k \geq 3$).

For the Oohara-Nakamura method, the source of the first Poisson equation is 
$\vec{\nabla}\cdot\vec{S}$~:~the degree of the harmonic is $l+1$ and the 
decrease is as $r^{-\left(k +1\right)}$. These two effects are opposed 
concerning the convergence properties established in Sec. \ref{converge}.
So, in the case no logarithm appear during the calculation to find $\chi$,
$\chi$ contains one spherical harmonic $l+1$ and decreases as 
$r^{-\left(k-1\right)}$ and so $\vec{\nabla} \chi$, part of the source 
of Eq. (\ref{eqoohara}), contains one spherical harmonic with $l+2$ and acting 
like $r^{-k}$ at infinity. So the conditions for the appearance of a 
Gibbs-like phenomenon are ``harder'' by two degrees than for a scalar Poisson
equation and occurs for a source with a spherical index $l+2$.

\subsubsection{The Shibata scheme}
Suppose we consider the same source than in the previous section. The 
convergence properties for the equation for $\vec{W}$ are the same than for 
a usual scalar Poisson equation.

Concerning the equation for $\chi$ the source is $-\vec{r}\cdot\vec{S}$. 
Performing such an operation on $\vec{S}$ increases 
the degree of the spherical harmonics by one unit. At the same time, the decrease of 
the source is slower, due to multiplication by $r$ everywhere. Those two 
phenomena have the same effect on the convergence criterion we previously 
established. As for the Oohara-Nakamura scheme, the criterion are ``harder'' by two 
degrees but the Gibbs-phenomenon occurs for a source in $l+1$.

\subsubsection {Convergence properties}

We are now able to deduce convergence properties for the two schemes. From 
the study above, we can see that if the condition for the appearance of 
the Gibbs-like phenomenon is the same, it is not associated with the same 
index $l$. This results in the two following properties~:

\begin{itemize}
\item{\bf Property 1 : }if the source of a vectorial Poisson equation 
is decreasing as $r^{-k}$ at infinity ( $k \geq 3$ for the Oohara-Nakamura scheme 
and $k \geq 4$ for the Shibata one)
and does not contain any spherical harmonics with $l\geq k-5$, then the 
error is evanescent.

\item{\bf Property 2 : }if the source decreases at least as $r^{-k}$ 
at infinity then 
the error is decreasing at least as $N^{-2\left(k-2\right)}$ for the 
Oohara-Nakamura method ($k\geq3$) and at least as $N^{-2\left(k-3\right)}$ for the 
Shibata one ($k \geq 4$).
\end{itemize}

\section{Accuracy of the vectorial Poisson equation solvers 
estimated by comparison with analytical solutions}\label{secvectconv}

To check the validity of the schemes and their convergence, we used the same 
method than for the scalar Poisson equation, that is the use of analytical 
solutions of various properties. The solutions associated 
with the sources have been obtained by following analytically the Shibata 
scheme.

\subsection {Continuous source}
Let us consider the case of a continuous source extending to infinity, say 
for example, in the external compactified domain, for $r>R$
\begin{equation}
\label{source_un}
S^x = \frac{x}{r^{n+5}} \phantom{mm} {\mathrm ;} \phantom{mm}
S^y = \frac{y}{r^{n+5}} \phantom{mm} {\mathrm ;} \phantom{mm}
S^z = \frac{z}{r^{n+5}}
\end{equation}
and for $r<R$
\begin{equation}
\label{source_deux}
S^x = \frac{x}{R^{n+5}} \phantom{mm} {\mathrm ;} \phantom{mm}
S^y = \frac{y}{R^{n+5}} \phantom{mm} {\mathrm ;} \phantom{mm}
S^z = \frac{z}{R^{n+5}}.
\end{equation}
Note that this source is $\cont{0}$, the minimum requirement for the Oohara-Nakamura
 method to be applicable.

\noindent For $n \not= 0$, the associated solution in the external domain is
\begin{equation}
N^x = \frac{1}{\left(\lambda +1\right) n \left(n+3\right)}
	\frac{x}{r^{n+3}}-\frac{n+5}{\left(\lambda+1\right)15n}
		\frac{x}{R^nr^3}
\end{equation}
and for $r<R$
\begin{equation}
N^x = \frac{1}{10 \left(\lambda+1\right)}\frac{xr^2}{R^{n+5}}
		- \frac{n+5}{\left(\lambda+1\right)6\left(n+3\right)}
			\frac{x}{R^{n+3}}.
\end{equation}
the other components being obtained by permutation of $x$, $y$ and $z$.

\begin{figure}
\centerline{\includegraphics[height=9cm,angle=-90]{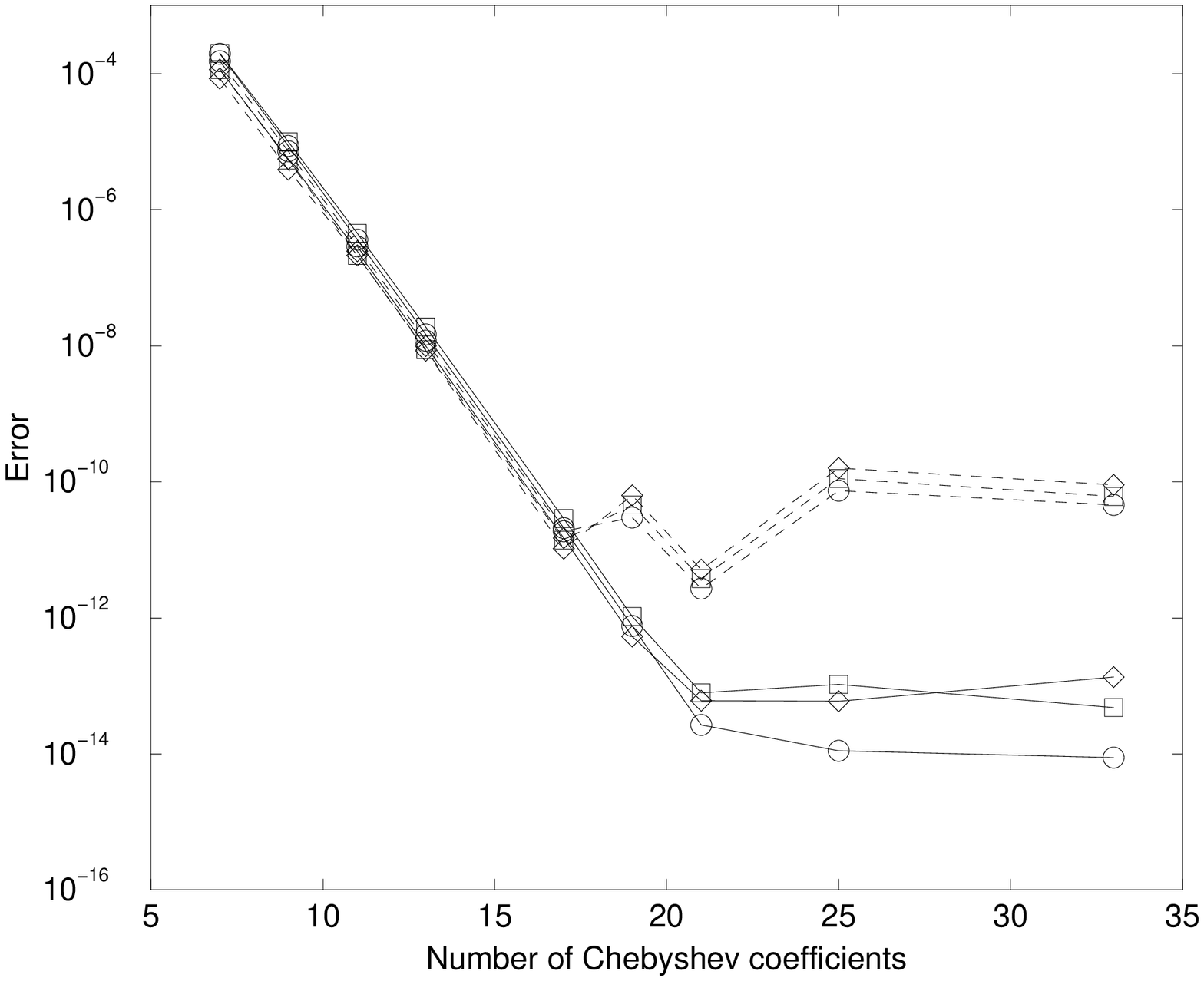}}
\vskip12pt
\caption
{\label{cont_1_z}
Error on the $z$ component for a continuous source extending to 
infinity (Eqs. \ref{source_un}-\ref{source_deux} with $n = 1$).

The scale for the number of coefficients is linear.

The solid lines represent the Shibata scheme and the dashed lines the 
Oohara-Nakamura one.

The circles represent the error in the kernel, the squares in the shell and 
the diamonds in the external domain.}
\end{figure}

For $n \not= 0$ no Gibbs-phenomenon occurs by solving the equations with 
$\vec{S}$ as source. For $n \leq 2$, a Gibbs-like phenomenon should appear due 
to the vectorial nature of Eq. (\ref{e:poisson_vect}). It is not the case because of 
simplifications due to the symmetry of the source. It just shows that the 
two convergence criteria established above are rather pessimistic. The 
evanescent error is shown in Fig. \ref{cont_1_z}. As for the scalar case, 
a saturation is attained at a level of approximatively $10^{-11}$.

\subsection {A vectorial Gibbs-like phenomenon}
At this point, we would like to exhibit an analytical solution that 
produces a Gibbs-phenomenon which arises from the vectorial nature 
of Eq. (\ref{e:poisson_vect}).
Let us consider the following source
\begin{equation}
\label{source_gibbs_vect}
S^z = \frac{z}{r^7}
\end{equation}
in the external compactified domain, and for $r<R$
\begin{equation}
S^z = \frac{z}{R^7}.
\end{equation}
We set the two other components to zero in all space.

If we solve the scalar Poisson equation with $S^z$ as source, the error 
will be evanescent, as shown by the study of Sec. \ref{converge}. But, according to 
the conclusion we obtained concerning the convergence criterion of a 
vectorial Poisson equation, a Gibbs-like phenomenon should appear
 due to the vectorial nature of Eq. (\ref{e:poisson_vect}).

\noindent In the external domain, the associated solution is

\begin{eqnarray}
N^x &=& -\frac{1}{2}\frac{\lambda}{\lambda+1} \left[
\frac{z^2x}{r^7}\left(-\frac{9}{14}+\ln\left(R\right)-\ln\left(r\right)\right)
+\frac{7}{10}\frac{z^2x}{r^5R^2}\right. \\
\nonumber
& & \left. +\frac{x}{r^5}\left(\frac{59}{350}+\frac{\ln\left(r\right)
		-\ln\left(R\right)}{5}\right) -\frac{7}{30}\frac{x}{r^3R^2}
\right] \\
\nonumber
N^y &=& -\frac{1}{2}\frac{\lambda}{\lambda+1} \left[
\frac{z^2y}{r^7}\left(-\frac{9}{14}+\ln\left(R\right)-\ln\left(r\right)\right)
+\frac{7}{10}\frac{z^2y}{r^5R^2}\right. \\
\nonumber
& & \left. +\frac{y}{r^5}\left(\frac{59}{350}+\frac{\ln\left(r\right)
		-\ln\left(R\right)}{5}\right) -\frac{7}{30}\frac{y}{r^3R^2}
\right] \\
\nonumber
N^z &=& \frac{1}{2}\frac{\lambda+2}{\lambda+1}z\left(\frac{1}{10r^5}
	-\frac{7}{30}\frac{1}{r^3R^2}\right)  
	-\frac{1}{2}\frac{\lambda}{\lambda+1}\left[\frac{z^3}{r^7}\left(
	-\frac{9}{14}+\ln\left(R\right)-\ln\left(r\right)\right) \right. \\
\nonumber
& & \left. + \frac{7}{10}\frac{z^3}{R^2r^5}+
\frac{z}{r^5}\left(\frac{3}{5}\left(\ln\left(r\right)-
\ln\left(R\right)\right)+\frac{71}{175}\right)-\frac{7}{15}\frac{z}{r^3R^2}
\right]
\end{eqnarray}
and for $r<R$, we found
\begin{eqnarray}
N^x &=& -\frac{1}{2}\frac{\lambda}{\lambda+1}x\left(\frac{2}{35}
\frac{z^2}{R^7}+\frac{1}{35}\frac{r^2}{R^7}-\frac{7}{75}\frac{1}{R^5}\right)
\\
\nonumber
N^y &=& -\frac{1}{2}\frac{\lambda}{\lambda+1}y\left(\frac{2}{35}
\frac{z^2}{R^7}+\frac{1}{35}\frac{r^2}{R^7}-\frac{7}{75}\frac{1}{R^5}\right)
\\
\nonumber
N^z &=& \frac{1}{2}\frac{\lambda+2}{\lambda+1}z\left(\frac{1}{10}
\frac{r^2}{R^7}-\frac{7}{30}\frac{1}{R^5}\right)
-\frac{1}{2}\frac{\lambda}{\lambda+1}z\left(\frac{2}{35}\frac{z^2}{R^7}
-\frac{1}{70}\frac{r^2}{R^7}-\frac{7}{150}\frac{1}{R^5}\right).
\end{eqnarray}

\begin{figure}
\centerline{\includegraphics[height=9cm,angle=-90]{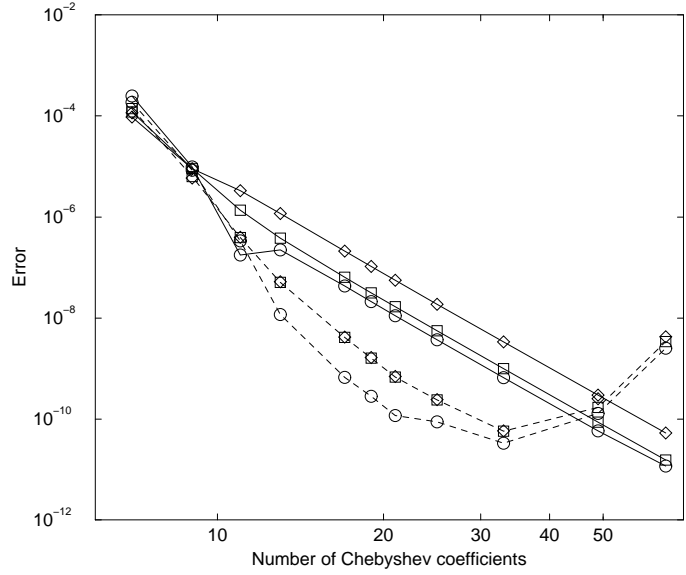}}
\vskip12pt
\caption{\label{gibbs_z}  
Error on the $z$ component for a source implying a Gibbs-like
phenomenon.

The scale for the number of coefficients is logarithmic.

The solid lines represent the Shibata scheme and the dashed lines the 
Oohara-Nakamura one.

The circles represent the error in the kernel, the squares in the shell and 
the diamonds in the external domain.}
\end{figure}

As expected, Fig. \ref{gibbs_z} shows an error obeying a power-law.
This feature is more evident in the external domain where the particular solution 
is directly present. The Gibbs-like phenomenon appears for the two schemes.
Let us apply Property 2 to determine the exponent of the power 
law. The source of the equation decreases as $r^{-6}$. This implies
that the error for the Oohara-Nakamura scheme should decrease at least as $N^{-8}$ 
and as $N^{-6}$ for the Shibata one. This is well confirmed for the Shibata 
scheme which exhibits an exponent $-6.4$. For the Oohara scheme it turns out that 
the criterium is rather pessimistic for the error decreases faster than $N^{-12}$.

\subsection{A discontinuous source}

As previously explained, the Oohara-Nakamura scheme fails to solve Eq. 
(\ref{e:poisson_vect}) in the case of a discontinuous source. We will now consider 
such a source and show that the Shibata method is efficient, even in such 
a case.

\noindent In the compactified domain, $r>R$, we choose the following solution
\begin{equation}
N^x = \frac{x}{r^n}
\end{equation}
For $r<R$, we ensure the continuity of the solution and its derivative by
choosing~:
\begin{equation}
N^x = x\left(ar^6+br^4\right),
\end{equation}
where $a = -\displaystyle\frac{4+n}{2R^{n+6}}$ and 
$b = \displaystyle\frac{6+n}{2R^{n+4}}$.
\vspace{0.3cm}
The associated source is obtained by calculating the left-hand side of 
Eq. (\ref{e:poisson_vect}). In the external domain we obtain
\begin{eqnarray}
\label{gibbs_un}
S^x &=& n\left(n-3-3\lambda\right)\frac{x}{r^{n+2}}+n\left(n+2\right)
\lambda\frac{x^3}{r^{n+4}} \\
\nonumber
S^y &=& -\lambda n \frac{y}{r^{n+2}}+n\left(n+2\right)\lambda\frac{x^2 y}
{r^{n+4}} \\
\nonumber
S^z &=& -\lambda n \frac{z}{r^{n+2}}+n\left(n+2\right)\lambda\frac{x^2 z}
{r^{n+4}}
\end{eqnarray}
and for $r<R$, we have
\begin{eqnarray}
\label{gibbs_deux}
S^x &=& x\left[\left(54+18\lambda\right)ar^4
+\left(28+12\lambda\right)br^2\right]+\lambda x^3\left(24ar^2+8b\right) \\
\nonumber
S^y &=& \lambda y\left[6ar^4+4br^2+x^2\left(24ar^2+8b\right)\right] \\
\nonumber
S^z &=& \lambda z\left[6ar^4+4br^2+x^2\left(24ar^2+8b\right)\right].
\end{eqnarray}

Depending on the value of $n$, the error is evanescent or not. Only a few 
spherical harmonics are present in the source and we can show that we 
expect, for example, an evanescent error for $n=4$ and a Gibbs phenomenon 
for $n=5$. This might seem not to be in agreement with the convergence criterion 
previously established, but the reader should recall that they are rather 
general and much more pessimistic to handle simple sources such as the ones 
which are considered here.

\begin{figure}
\centerline{\includegraphics[height=9cm,angle=-90]{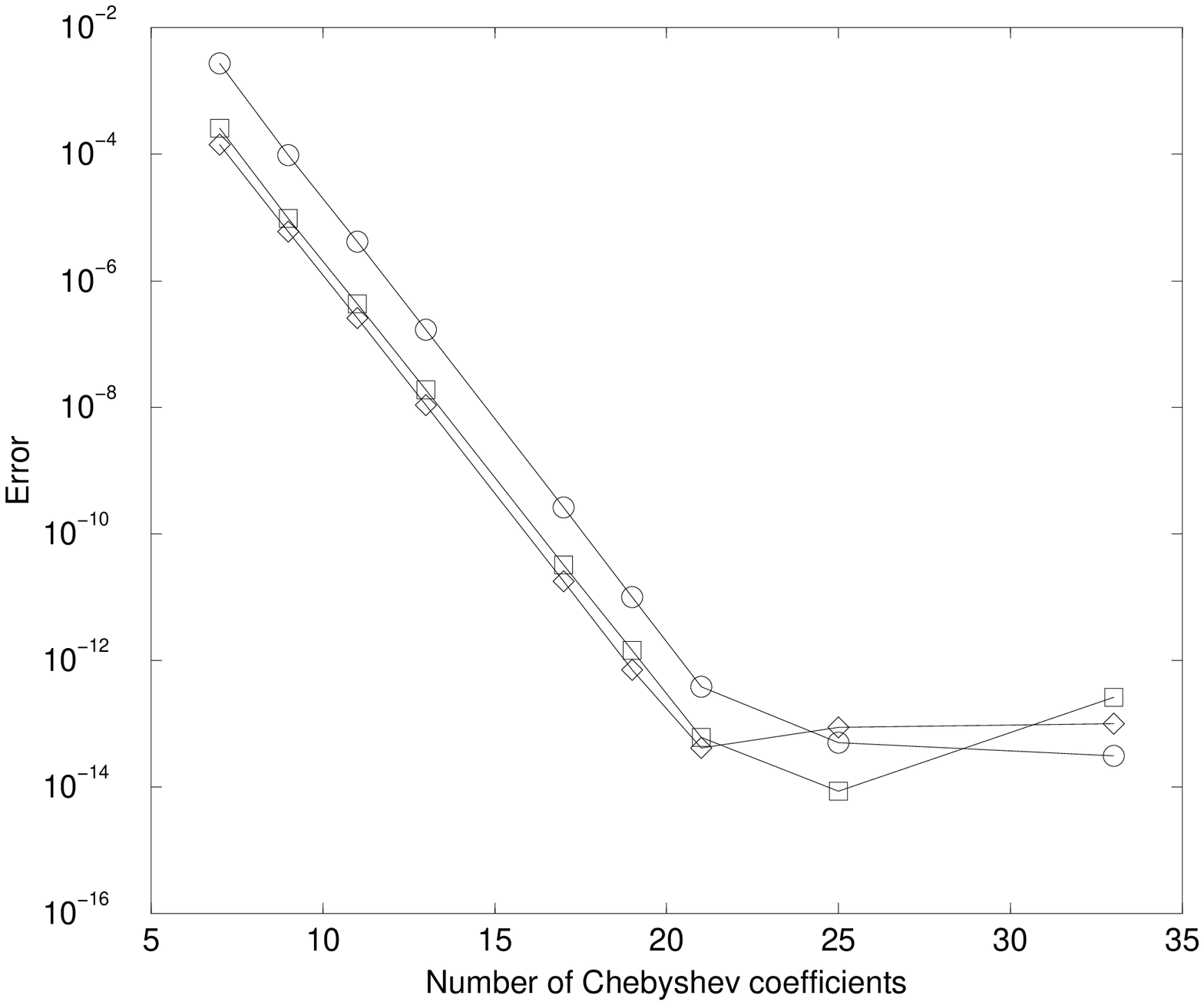}}
\vskip12pt
\caption{\label{inf_x_4} 

Error on the $x$ component for discontinuous source (Eqs. \ref{gibbs_un}-
\ref{gibbs_deux} with $n=4$).

The scale for the number of coefficients is linear.

The circles represent the error in the kernel, the squares in the shell and 
the diamonds in the external domain.}
\end{figure}

\begin{figure}
\centerline{\includegraphics[height=9cm,angle=-90]{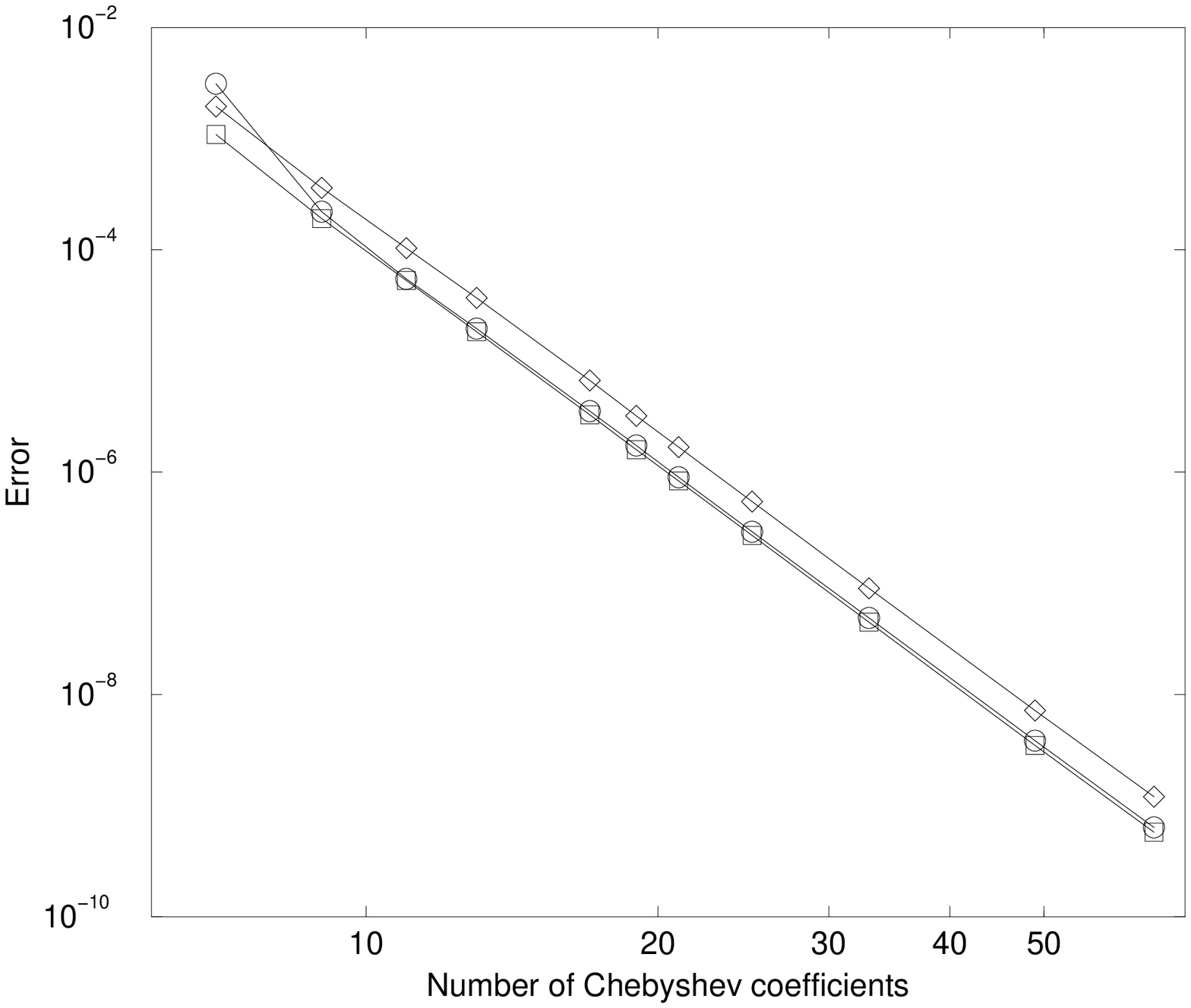}}
\vskip12pt
\caption{\label{inf_x_5}
Same as Fig. \ref{inf_x_4} but for $n=5$ ; the scale for the number of coefficients 
is now logarithmic.}
\end{figure}

The results presented by Figs. \ref{inf_x_4} and \ref{inf_x_5} show that 
the discontinuity of the source has no effect on the resolution of the 
vectorial Poisson equation, as long as the Shibata scheme is used. For 
$n=5$, the source is like $r^{-6}$ at infinity and we expect an error 
decreasing more rapidly than $N^{-6}$. Fig. 
\ref{inf_x_5} shows an extremely good agreement with the prediction, for the 
power-law exhibits an exponent of $-6.4$.

\section {Developments}\label{sec_devel}
In this section we present some extension of this work, that is solving 
more complicated equations using the schemes presented here as milestones.

The first extension that has been conducted regards non-spherical 
domains, with spheroidal shapes (i.e. they must have the same topology 
as a sphere). This is very useful for we can define the boundary of each domain 
to match with surfaces of discontinuity, like stellar surfaces, so that each 
field is $\cont{\infty}$ in each domain preventing any Gibbs phenomenon.
Thanks to some mapping onto a sphere, solving the Poisson equations with such boundaries reduces to the 
spherical case, with correction terms appearing in the source. The equation 
is then solved by iteration.
The method is described in details in \cite{BonazGM98}. In that paper 
the calculation of the Mac-Laurin and of the Roche ellipsoids have been 
compared with the analytical solutions. The behavior of the error when one 
increases the number of coefficients happens to be evanescent (see Figs.
5 and 6 of Ref.\cite{BonazGM98}). Those calculations being done in the Newtonian case, 
all the sources are compactly supported. This shows that the non-sphericity 
does not introduce any new Gibbs phenomenon with respect to the spherical 
case.

Concerning calculations in general relativity (i.e. with sources 
extending to infinity), results have been obtained for rapidly rotating 
strange stars in \cite{GourgHLPBM99} using non-spherical domains. Convergence 
properties have not been fully explored for there exist no analytical 
solution to compare with. Anyway we can suppose that, the sources containing 
almost every spherical harmonics, the convergence will no longer be 
evanescent but rather follows a power-law.

\begin{figure}
\centerline{\includegraphics[height=9cm,angle=-90]{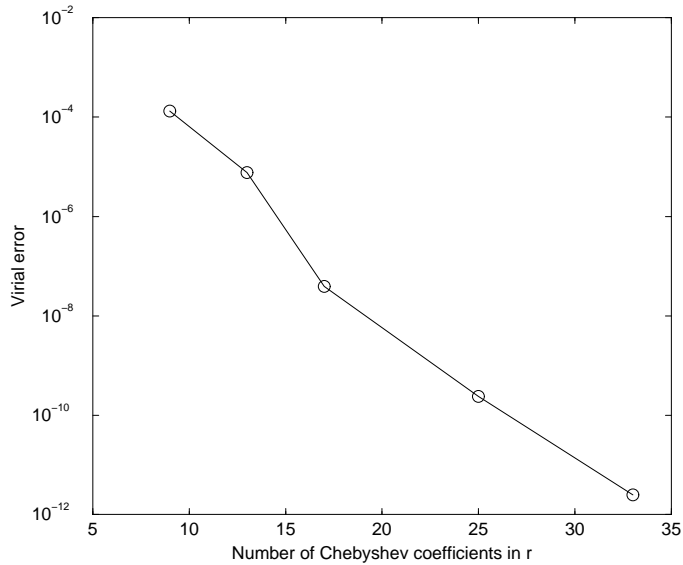}}
\vskip12pt
\caption{\label{converge_deux}
Relative error, estimated by means of the virial theorem, for a Newtonian 
irrotational binary star calculation with respect to the number 
of Chebyshev coefficients.}
\end{figure}

Another important extension of this work is to deal with two bodies, for 
example orbiting binary neutron stars. This case has been successfully 
studied in \cite{BonazGM99a, GourgGTMB01}, by means of the Poisson solvers 
presented here. The main difference with the cases we 
discussed in the present paper is that the sources are no longer spheroidal 
but are concentrated on two spheroidal domains, being the two 
stars. An equation of the type (\ref{e:poisson_scal}) is then split into two 
parts~:
\begin{eqnarray}
\Delta F_1 &=& S_1 \\
\nonumber \Delta F_2 &=& S_2,
\end{eqnarray}
where the real source is $S = S_1 + S_2$.
We use two sets of spherical coordinates, one centered on each star and 
the splitting is done so that $S_1$ is mainly centered on the first star 
and $S_2$ 
on the other one (see \cite{GourgGTMB01} for details). The sources $S_i$ are then well described in spheroidal 
topology and the total equation is well solved, the solution being 
$F = F_1 + F_2$. We used that method to compute Newtonian configurations 
and compare them with semi-analytical solutions.
Fig. \ref{converge_deux} shows the error done for the same configuration as
Fig. 7 of \cite{GourgGTMB01} for a coordinate separation of $100$ km. This 
calculation being Newtonian, the sources are compactly supported and the 
error seems to be evanescent, but we have to be cautious for the number 
of coefficients of the expansion are not maintained fixed. Extensive 
convergence properties have not been conducted but it seems that the splitting 
of the equation into two parts does not introduce any new Gibbs phenomenon. 
As for the single body problem, convergence of calculations with sources extending to 
infinity (i.e. in general relativity), have not been studied but we 
expect a Gibbs phenomenon to occur for the sources contain almost every 
spherical harmonics.

To finish with the extension of this work let us mention the case of 
black holes. In that case the equations are not solve in all space but only 
on the domain exterior to the holes horizons. This means that we have to 
remove the kernel from the computational domain. The regularity condition
at the origin is then replaced by a boundary condition on the boundary 
of the innermost shell. We have been able to use that to impose condition 
on the value of the solution (Dirichlet problem) or on its first radial 
derivative (Neumann problem). This extension has nothing to do with the 
compactified domain and we expect the convergence properties to be the 
same as the one exhibited in the present work. We are currently applying 
this  to compute realistic physical binary black holes configurations.

\section{Conclusion}

We have presented a scalar Poisson equation solver based on spectral method. 
It enables us to solve the Poisson equation for a source extending to 
infinity and going to zero at least like $r^{-3}$. Our multi-domain 
approach enables to deal with a source which is $\cont{\infty}$ in
each domain. Nevertheless some Gibbs phenomenon can appear due to the 
existence of particular solutions which contain logarithm functions in the 
external domain. Such functions are not well described in terms of Chebyshev
polynomials, resulting in a Gibbs-like phenomenon. We exhibited the conditions
for the appearance of such an effect and quantified it, leading to the 
conclusion that, for a source decaying as $r^{-k}$ ($k \geq
3$), the error of the numerical solution is evanescent if the source
does not contain any spherical harmonics with index $l\geq k-3$.
Otherwise, the error decreases at least as
$N^{-2(k-2)}$, $N$ being the number of Chebyshev coefficients.

We used this scalar Poisson equation solver to solve the generalized 
vectorial Poisson 
equation given by Eq. (\ref{e:poisson_vect}) for a source going to zero at least like 
$r^{-4}$. Three different schemes have been discussed.
We showed than the one proposed by Bowen \& York \cite{BowenY80} is
not applicable to domains extending up to infinity, by means of our methods,
because it gives rise to diverging 
auxiliary quantities. The scheme from Oohara \& Nakamura 
\cite{OoharN97} is applicable as long as the source is continuous and has 
been successfully implemented. 
The last scheme, proposed by Oohara, Nakamura and Shibata \cite{OoharNS97},
is applicable even for discontinuous sources and has been successfully 
implemented too. The 
convergence properties of the two implemented schemes have been derived from 
the ones of the scalar Poisson equation solver and checked by comparison 
between calculated and analytical solutions.

\end{article}
\end{document}